\documentclass[%
reprint,
superscriptaddress,
 amsmath,amssymb,
 aps,
 pra,
]{revtex4-2}

\usepackage{dcolumn}
\usepackage{bm}
\usepackage{graphicx} 

\begin{document}
 
\preprint{APS/123-QED}

\title{Transition from exceptional points to observable nonlinear bifurcation points \\in anti-PT symmetric coupled cavity systems}

\author{Takahiro Uemura}
\affiliation{Department of Physics, Institute of Science Tokyo, 2-12-1 Ookayama, Meguro-ku, Tokyo 152-8551, Japan}
\affiliation{NTT Basic Research Laboratories, NTT Inc., 3-1 Morinosato-Wakamiya, Atsugi-shi, Kanagawa 243-0198, Japan}

\author{Kenta Takata}
\affiliation{NTT Basic Research Laboratories, NTT Inc., 3-1 Morinosato-Wakamiya, Atsugi-shi, Kanagawa 243-0198, Japan}
\affiliation{Nanophotonics Center, NTT Inc., 3-1, Morinosato-Wakamiya, Atsugi-shi, Kanagawa 243-0198, Japan}

\author{Masaya Notomi}
\affiliation{Department of Physics, Institute of Science Tokyo, 2-12-1 Ookayama, Meguro-ku, Tokyo 152-8551, Japan}
\affiliation{NTT Basic Research Laboratories, NTT Inc., 3-1 Morinosato-Wakamiya, Atsugi-shi, Kanagawa 243-0198, Japan}
\affiliation{Nanophotonics Center, NTT Inc., 3-1, Morinosato-Wakamiya, Atsugi-shi, Kanagawa 243-0198, Japan}
\email{notomi@phys.sci.isct.ac.jp}

\begin{abstract}
  Exceptional points (EPs) in anti-parity-time (APT)-symmetric systems have attracted significant interest. While linear APT-symmetric systems exhibit structural similarities with nonlinear dissipative systems, such as mutually injection-locked lasers, the correspondence between exceptional points in linear non-Hermitian Hamiltonians and bifurcation phenomena in nonlinear lasing dynamics has remained unclear. We demonstrated that, in a two-cavity system with APT symmetry and gain saturation nonlinearity, an EP coincides with a bifurcation point of nonlinear equilibrium states, which appears exactly at the lasing threshold. Although the EP and the bifurcation point originate from fundamentally different physical concepts, the bifurcation point is observable and retains key EP characteristics even above the lasing threshold. Notably, the bifurcation point that originates from the linear EP also bridges linear and nonlinear dynamics of the system: it serves as an accessible transition point in the nonlinear dynamics between the limit-cycle and synchronization regimes. Furthermore, we clarified that beat oscillation that conserves the energy difference, which is a unique dynamic in the weak-coupling regime of a linear APT system, evolves into a nonlinear limit cycle with equal amplitudes in the two cavities in the presence of gain saturation. Our findings establish a direct link between EP-induced bifurcation points and nonlinear dynamics, providing fundamental insights into non-Hermitian and nonlinear optical systems.
\end{abstract}

\maketitle

\section{Introduction}

Parity-time (PT) and anti-parity-time (APT) symmetric systems have received considerable attention in recent years due to their distinctive physical characteristics \cite{PhysRevLett.80.5243, Bender_2007, Regensburger2012, Feng2017, Ozdemir2019, doi:10.1126/science.aar7709, Choi2018, Zhang2020, Li2024}. Exceptional points (EPs) in PT- and APT-symmetric systems are parameter points at which two eigenvalues and their corresponding eigenvectors simultaneously coalesce \cite{Heiss2001}. In linear open systems, PT-symmetric phases undergo a phase transition at EPs, which gives rise to various intriguing phenomena, such as superluminal transmission \cite{PhysRevApplied.7.054023}, non-reciprocal transmission \cite{Peng2014, doi:10.1073/pnas.1603318113}, and enhanced sensing capabilities \cite{Park2020, Duggan2022, Kononchuk2022}.

While EPs have been extensively studied in linear gain-loss systems, these studies typically focus on the non-lasing states, where the system remains below threshold and can be well described by linear eigenvalue analysis \cite{Takata:21}. In contrast, beyond the lasing threshold, nonlinear effects—particularly gain saturation—significantly influence the stability of oscillatory states. The physics in this regime is governed by nonlinear dynamics, characterized by a rich variety of phenomena such as limit cycles, synchronization, and chaos \cite{strogatz2018nonlinear}, which strongly contrasts with the linear regime traditionally associated with non-Hermitian optics and EPs. Furthermore, nonlinear open systems exhibit qualitative changes in their dynamics across bifurcation points, which function as phase transition points of the nonlinear dynamics, similar to EPs in linear open systems. Even though gain saturation nonlinearity is common in optical systems, it is not fully understood how the properties of linear EPs are carried over into the nonlinear dynamics of the system. While several studies have investigated EPs in nonlinear systems—with approaches ranging from identifying them with bifurcation points \cite{10.1093/nsr/nwac259, PhysRevLett.129.013901, PhysRevLett.130.266901}, to defining them via the defectiveness of the Jacobian matrix at steady states \cite{Roy:21}, or characterizing them by the emergence of unidirectional coupling \cite{Longhi:17}—these interpretations remain inconsistent and do not provide a direct correspondence with the original concept of EPs in linear non-Hermitian systems. In contrast, the formulation adopted in Refs. \cite{Ji2023, Fischer2024}, which defines nonlinear EPs as bifurcation points at which the nonlinear Hamiltonian itself becomes defective, offers a natural extension of the linear EP concept to the nonlinear regime. However, as discussed in detail in Ref. \cite{Ji2023}, the EPs in PT-symmetric coupled cavity systems under nonlinear conditions are not directly observable, since the equilibrium states become unstable due to a mismatch in the time scales between carrier dynamics and the cavity decay rate. In other words, it remains unclear how the characteristics of linear EPs are reflected in experimentally observable dynamics upon the inclusion of nonlinearities. This raises a fundamental question: can one define nonlinear EPs that both preserve their unique features in the linear regime and manifest in physically accessible, observable dynamics? 

APT-symmetric coupled cavity systems offer a promising platform in this context, as their coupling schemes closely resemble those of injection-locked lasers. The characteristics of nonlinear bifurcations and nonlinear dynamics emerging in injection-locked systems have been extensively studied in various contexts, including injection-locked lasers \cite{siegman1986lasers, 1071632, 1071166, 1070479, 62107}, optical parametric oscillators (OPOs) \cite{PhysRevA.92.043821}, and coherent Ising machines (CIMs) \cite{PhysRevA.88.063853, doi:10.1126/science.aah4243}. Therefore, analyzing the properties of APT-symmetric systems in the nonlinear regime serves as a bridge between the characteristics of linear EPs and well-established nonlinear physics. Furthermore, recent studies on injection-locked laser systems have shown that frequency synchronization emerges at bifurcation points analogous to eigenvalue coalescence at EPs in APT-symmetric systems~\cite{Takemura2021}, which suggests that nonlinear bifurcations can inherit key features of linear EPs. Nevertheless, APT-symmetric systems are no exception to the broader trend in non-Hermitian photonics, where most studies focus on solving linear eigenvalue problems while neglecting gain saturation and other nonlinear contributions. For instance, APT-symmetric configurations have demonstrated observable EP-related phenomena in the linear regime~\cite{doi:10.1126/sciadv.adr8283}, yet their behavior in the nonlinear regime remains unknown. Conversely, most models of injection-locked lasers rely on rate equations or reduced-order approximations (e.g., phase-reduction methods~\cite{Nakao02042016}), emphasizing synchronization and chaotic dynamics without addressing their relationship to linear EP behavior~\cite{PhysRevA.65.063815, PhysRevLett.65.1575, PhysRevE.55.3865}.

This study investigates the relationship between linear EPs and nonlinear bifurcations in the simplest APT-symmetric system—two coupled cavities with imaginary coupling—and reveals that, precisely at the lasing threshold, a nonlinear bifurcation point emerges that corresponds to the EP of the associated linear APT-symmetric system. Although the EP and the bifurcation point are described by fundamentally different physical equations, our results demonstrate that the linear EP transforms into a stable nonlinear bifurcation point, inheriting many of its key properties. Moreover, the EP-derived bifurcation remains observable, manifesting as `symmetric' and `broken' solution branches that are directly inherited from the phases of the linear system. We also demonstrate that these conclusions remain robust even when including carrier dynamics, confirming that these phenomena are indeed observable in realistic physical systems: we clarify that the energy-difference-conserving beat oscillations characteristic of the linear APT-symmetric system transform into a nonlinear limit cycle, showing that the nonlinear EP emerges precisely at the infinite-period limit of this cycle. Our results elucidate how various physical features of the linear APT-symmetric system manifest as distinct phenomena in the nonlinear regime, thereby clarifying the connection between linear EPs and nonlinear bifurcation points—a stark contrast to prior work focusing on the trajectories of unobservable bifurcation points.

\section{Eigenstates and equilibrium states of the nonlinear Hamiltonian}

\subsection{Equations of motion}

\begin{figure}%
    \includegraphics[width=7cm]{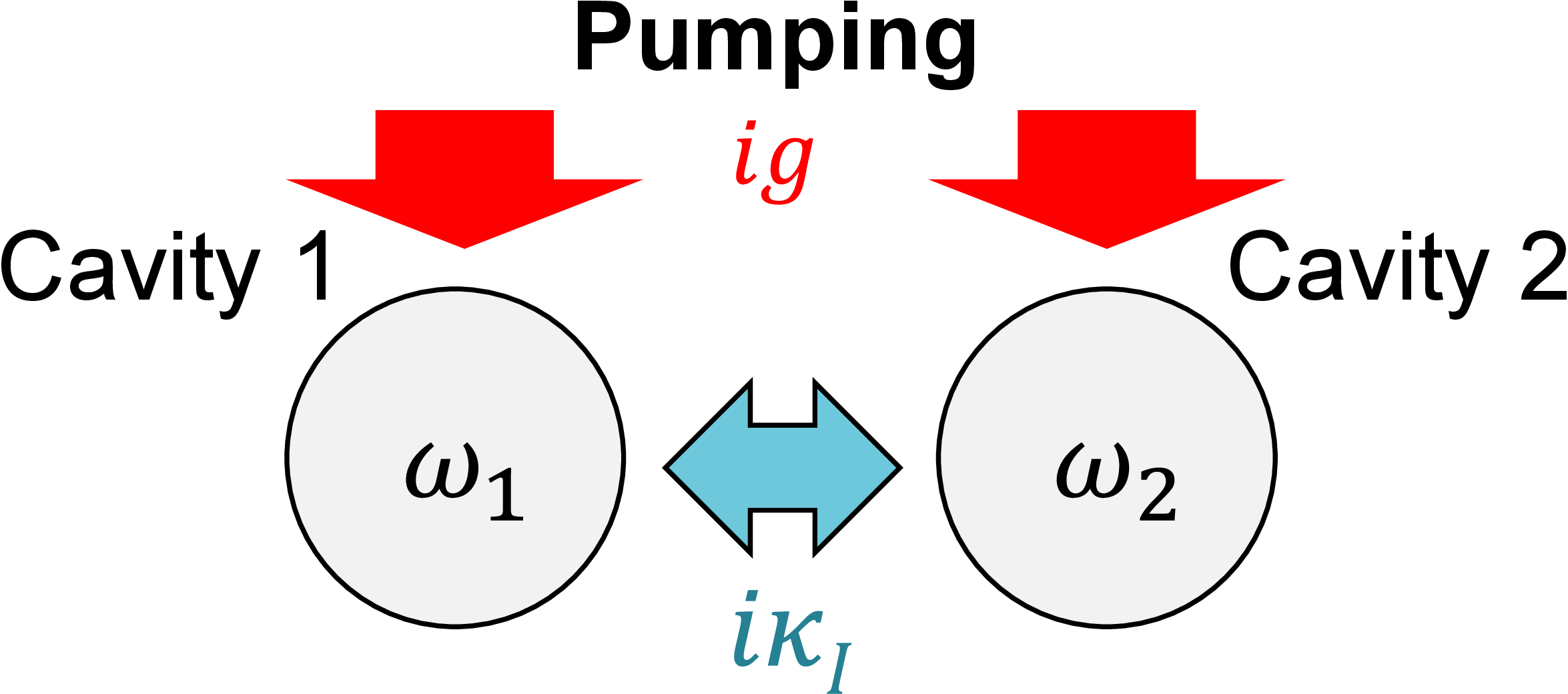}
    \caption{\label{fig_schematic}
      Schematic of the coupled cavities system. The two cavities are dissipatively coupled with a coupling rate $\kappa_I$ and have a frequency detuning $\Delta \omega := \omega_1 - \omega_2 $. The gain is applied to both cavities with a gain rate $g$. The gain saturation effect is modeled by the nonlinear term $f(a_j) = \beta{|a_j|}^2$.
    }
\end{figure}
Our model is based on two dissipatively coupled optical cavities with gain, as shown in Fig. \ref{fig_schematic}. We assume that the two cavities are coupled with a pure imaginary coupling rate $\kappa_I$. This system can be realized by methods such as adjusting the coupling phase of the microring resonator via the waveguides \cite{Naweed:15}, or utilizing adiabatic elimination via cold cavities \cite{Takemura2021}. 
We describe the equations of motion using a coupled-mode theory with simplified gain saturation:
\begin{align}
  H(\boldsymbol{a}) &= \begin{bmatrix}
  \omega_1 + i(\gamma - \beta{|a_1|}^2) & -i\kappa_I \\
  -i\kappa_I & \omega_2+i(\gamma - \beta{|a_2|}^2) \\
  \end{bmatrix},
  \nonumber \\
  \label{eq:Hamiltonian}
  &\frac{\mathrm{d} }{\mathrm{d} t} \boldsymbol{a}(t) = -iH(\boldsymbol{a}(t)) \boldsymbol{a}(t)
\end{align}
Here, $\omega_1, \omega_2$ are the eigenfrequencies of the two cavities, $\gamma = g - |\kappa_I|$ represents the net gain, where $g$ represents the linear gain and the term $-|\kappa_I|$ represents the energy loss due to dissipative coupling. The term $\beta{|a_j|}^2$ represents the Stuart-Landau nonlinearity, where we consider only the lowest-order term related to gain saturation. For simplicity, we assume $\beta \in \mathbf{R}$, which implies that we neglect the term corresponding to the linewidth enhancement factor, as the focus is on understanding how the nonlinear term affects the behavior near the EPs. This assumption is justified in our anti-PT-symmetric optical system, where the gain applied to both cavities is equal and the real part of the net detuning is canceled out, making this approximation valid.

\subsection{Linear eigenvalue and eigenstates}

\begin{figure}%
  \includegraphics[width=8cm]{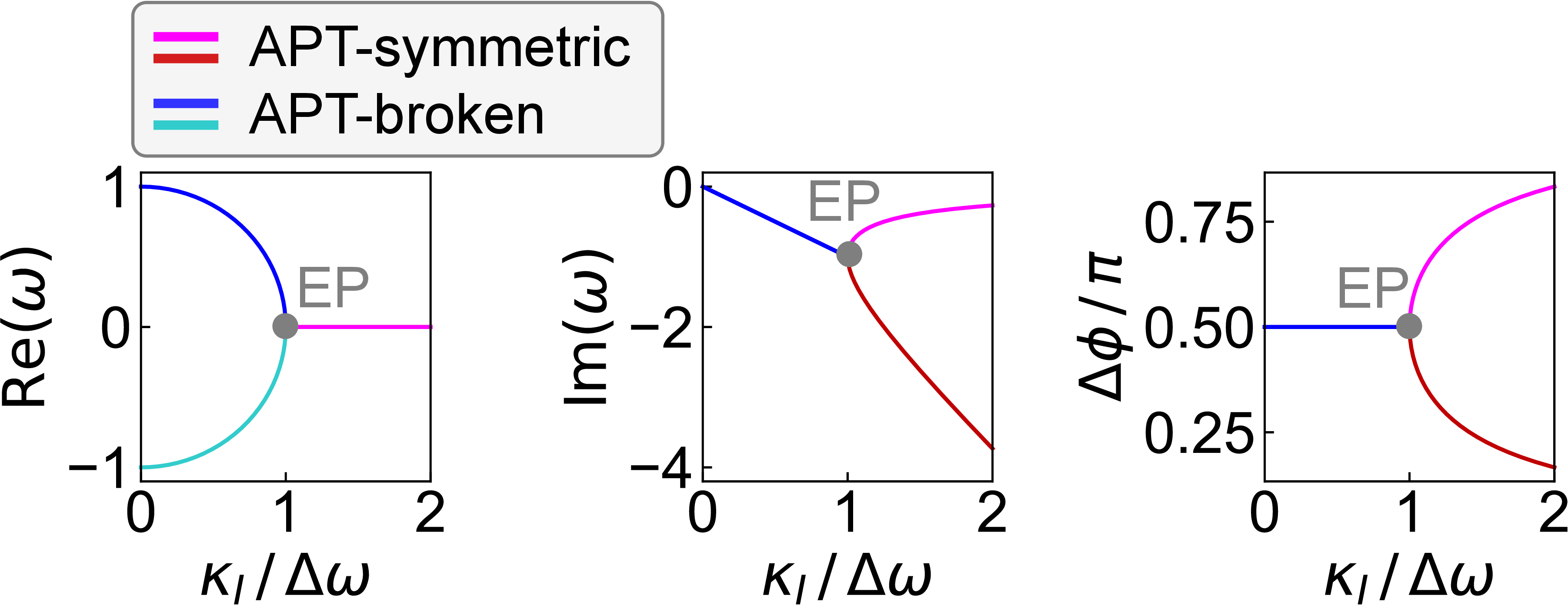}
  \caption{\label{fig_eigenstates}
    Complex eigenvalues of the Hamiltonian (Eq. \eqref{eq:Hamiltonian}) for the linear regime, $\beta=0$ as a function of the coupling strength $\kappa_I$ for $g=0$. The horizontal axis is normalized by $\Delta \omega$.
  }
\end{figure}

The system reduces to a linear eigenvalue problem when we set $\beta=0$. The linearized Hamiltonian $H_L$ can be expressed as:
\begin{align}
  \label{eq:Hamiltonian_linear}
  H_L 
  &= (\omega_0+i\gamma)I_2 + H_{L1}, \quad
  H_{L1} =
  \begin{bmatrix}
  \Delta \omega & -i\kappa_I \\
  -i\kappa_I & -\Delta \omega \\
  \end{bmatrix} 
\end{align}
Here, $\omega_0 = (\omega_1 + \omega_2)/2$ is the average frequency of the two cavities, $H_{L1}$ is the coupling matrix, $\Delta \omega = (\omega_1 - \omega_2) / 2$ is the frequency detuning between the two cavities, and $I_2$ is the identity matrix. 
The $H_{L1}$ matrix is anti-PT symmetric, meaning that it is invariant under the simultaneous transformation of parity and time reversal.
\begin{align}
  (PT) H_{L1} (PT)^{-1} = -H_{L1}, \nonumber\\
  P = \begin{bmatrix}
  0 & 1 \\
  1 & 0 \\
  \end{bmatrix}, \quad
  T = \mathcal{K}
\end{align}
where $\mathcal{K}$ is the complex conjugation operator. The eigenvalue problem of $H_L$ can be solved by diagonalizing the Hamiltonian matrix.
\begin{align}
\label{eq:eigenvalue}
  \omega = \omega_0 + i\gamma \pm \sqrt{ (\Delta \omega)^2 - \kappa_I^2 }
\end{align}
The Hamiltonian $H_L$ is defective at the EP, when $\kappa_I = \Delta \omega$ with the eigenvector $\boldsymbol{v}_\mathrm{EP} = [1, -i]^\top$. 
Moreover, the eigenstates for $\kappa_I < \Delta \omega$ and $\kappa_I > \Delta \omega$ corresponds to the APT-broken phase and APT-symmetric phase, respectively. The eigenvectors in these two phases $\boldsymbol{v}_\mathrm{bro}$ and $\boldsymbol{v}_\mathrm{sym}$ are given by:
\begin{align}
  \label{eq:eigenvector_broken}
  \boldsymbol{v}_\mathrm{bro} &= \begin{bmatrix}
  1 \\
  e^{-i\Delta \phi} \frac{\Delta \omega \mp \sqrt{ (\Delta \omega)^2 - \kappa_I^2 }}{\kappa_I} \\
  \end{bmatrix}, \quad \Delta \phi=\frac{\pi}{2}\\
  \label{eq:eigenvector_symmetric}
  \boldsymbol{v}_\mathrm{sym} &= \begin{bmatrix}
  1 \\
  e^{-i\Delta \phi} \\
  \end{bmatrix}, \quad
  \Delta \phi = \tan^{-1} \left( \frac{\pm \Delta \omega}{\sqrt{ \kappa_I^2 - (\Delta \omega)^2 }} \right)
\end{align}
The Hamiltonian $H_L$ is defective at the EP, occurring at $\kappa_I = \Delta \omega$ with the coalesced eigenvector $\boldsymbol{v}_{\mathrm{EP}} = [1, -i]^\top$. For $\kappa_I > \Delta \omega$, the eigenvectors $\boldsymbol{v}_\mathrm{sym}$ are invariant under the combined anti-parity-time (APT) operation, and this regime is referred to as the APT-symmetric phase. 
In contrast, for $\kappa_I < \Delta \omega$, the eigenvectors $\boldsymbol{v}_\mathrm{bro}$ no longer respect the APT symmetry—this is the APT-broken phase, where the APT symmetry is spontaneously broken in the eigenstates despite the Hamiltonian itself being APT-symmetric.
The key difference between $\boldsymbol{v}_{\mathrm{sym}}$ and $\boldsymbol{v}_{\mathrm{bro}}$ lies in the amplitude balance between the two cavities. In the APT-symmetric phase, the eigenvector satisfies $|\left(\boldsymbol{v}_{\mathrm{sym}}\right)_1| = |\left(\boldsymbol{v}_{\mathrm{sym}}\right)_2|$, indicating equal amplitude distribution. In contrast, the APT-broken phase exhibits an imbalance in the amplitudes, such that $|\left(\boldsymbol{v}_{\mathrm{bro}}\right)_1| \neq |\left(\boldsymbol{v}_{\mathrm{bro}}\right)_2|$.

Fig. \ref{fig_eigenstates} shows the complex eigenvalues and the eigenvector phase difference between the two cavities as a function of $\kappa_I$ for $g=0$. An EP appears at $\kappa_I = \Delta \omega$, which divides the system into the APT-symmetric phase $(\kappa_I > \Delta \omega)$ and the APT-broken phase $(\kappa_I < \Delta \omega)$. Focusing on the imaginary part of the eigenfrequencies, we observe that in the APT-broken phase, $\mathrm{Im}(\omega)$ increases as $\kappa_I$ increases. Moreover, once the system crosses the EP into the APT-symmetric phase, the imaginary part of the eigenfrequencies splits, and the imaginary part of one of the eigenfrequencies begins to decrease. Consequently, focusing on the upper branch, the EP corresponds to a dip in the imaginary part of the eigenfrequency, which indicates that the lasing threshold reaches its maximum at the EP.

When a linear gain $g > 0$ is introduced in the linear system (i.e., $\beta = 0$), lasing occurs at the point where $\mathrm{Im}(\omega) = 0$, which corresponds to the threshold condition. Beyond this threshold, $\mathrm{Im}(\omega)$ becomes positive, indicating an exponential growth of the field amplitudes. However, in practice, the amplitude does not diverge indefinitely due to the presence of gain saturation described in the next section.

\subsection{Nonlinear equilibrium states}

Next, let us consider the equilibrium states within the nonlinear framework, which is completely different from the linear eigenvalue problem.
In contrast to the linear regime, where gain and loss of the modes are described by the imaginary part of the eigenfrequency ($\omega \in \mathbf{C}$), we focus here on steady-state solutions in the nonlinear regime and therefore assume that $\omega \in \mathbf{R}$. This assumption is justified because we are interested in steady states without exponential growth or decay. The effect of gain is no longer treated as an imaginary frequency shift but is instead incorporated through a nonlinear saturation term that depends on the amplitude of the field. Although it is also possible to model amplification using a complex eigenfrequency $\omega = \omega_r + i \omega_i \in \mathbf{C}$, this approach is effectively equivalent to including a net gain term $\gamma$ on the right-hand side of the equations of motion. Therefore, the formulation under $\omega \in \mathbf{R}$ is sufficient for our analysis. By splitting the real and imaginary parts as $a_i = r_i e^{-i(\omega t + \phi_i)} \ (i=1,2)$ in \eqref{eq:Hamiltonian} and defining $\Delta \phi = \phi_2 - \phi_1$, we obtain  
\begin{align}
  \label{eq:nonlinear_r1}
  \frac{\mathrm{d} }{\mathrm{d} t} r_1 &= (\gamma - \beta r_1^2) r_1 + r_2 \kappa_I \cos \Delta \phi \\
  \label{eq:nonlinear_r2}
  \frac{\mathrm{d} }{\mathrm{d} t} r_2 &= (\gamma - \beta r_2^2) r_2 + r_1 \kappa_I \cos \Delta \phi \\
  \label{eq:nonlinear_phi1}
  \frac{\mathrm{d} }{\mathrm{d} t} \phi_1 &= -\omega + \omega_1 + \frac{r_2}{r_1} \kappa_I \sin \Delta \phi \\
  \label{eq:nonlinear_phi2}
  \frac{\mathrm{d} }{\mathrm{d} t} \phi_2 &= -\omega + \omega_2 - \frac{r_1}{r_2} \kappa_I \sin \Delta \phi
\end{align}
In the equilibrium states ($\mathrm{d} r_i / \mathrm{d} t  = 0$, $\mathrm{d} \phi_i / \mathrm{d} t = 0$), we obtain the following equations:  
\begin{align}
  \label{eq:steady1}
  \gamma(I_1 + I_2) - \beta(I_1^2 + I_2^2) + 2 \sqrt{I_1 I_2} \kappa_I \cos \Delta \phi &= 0 \\
  \label{eq:steady2}
  \gamma(I_1 - I_2) - \beta(I_1^2 - I_2^2)  &= 0 \\
  \label{eq:steady3}
  2\Delta \omega \sqrt{I_1 I_2} + (I_1 + I_2) \kappa_I \sin \Delta \phi &= 0
\end{align}
Where $I_i = r_i^2$. 
The frequency $\omega$ can be obtained from
\begin{align}
  \label{eq:frequency}
  \omega = \omega_0-\frac{(I_1 - I_2) \kappa_I \sin \Delta \phi}{2\sqrt{I_1 I_2}}
\end{align}
From \eqref{eq:steady2}, $I_1 = I_2$ or $\gamma - \beta(I_1 + I_2) = 0$ holds. The former solutions 
\begin{align}
  \label{eq:sol_symmetric}
  I_1 = I_2 = \frac{\gamma \pm \sqrt{\kappa_I^2 - {(\Delta \omega)}^2 }}{\beta}, \quad \omega = \omega_0
\end{align}
exhibit equal amplitudes between the two cavities ($I_1 = I_2$), a feature also seen in the APT-symmetric phase of the linear regime in \eqref{eq:eigenvector_symmetric}.
In contrast, the latter solutions
\begin{align}
  I_1 &= \frac{\gamma}{2\beta} \left(
    1 \pm \sqrt{ 1 - \frac{4\kappa_I^2}{ 4 {(\Delta \omega)}^2 + \gamma^2 } }
  \right) , \nonumber \\
  \label{eq:sol_broken}
  I_2 &= \frac{\gamma}{2\beta} \left(
    1 \mp \sqrt{ 1 - \frac{4\kappa_I^2}{ 4 {(\Delta \omega)}^2 + \gamma^2 } }
  \right)
\end{align}
with 
\begin{align}
  \label{eq:frequency_broken}
  \omega &= \omega_0 \pm \sqrt{ (\Delta \omega)^2 - \frac{\kappa_I^2}{ {1 + {\left( \frac{\gamma}{2\Delta \omega} \right)}^2 } } }
\end{align}
has different amplitudes between the two cavities ($I_1 \neq I_2$), which is a characteristic of the APT-broken phase in the linear regime \eqref{eq:eigenvector_broken}. We refer to these two types of solutions as the ``symmetric'' and ``broken'' states, respectively. 
The similarity in symmetry between the linear eigenstates and nonlinear equilibrium states is not coincidental, but rather a consequence of the underlying symmetry of the Hamiltonian.
The gain saturation terms in the nonlinear Hamiltonian, as defined in Eq.~\eqref{eq:Hamiltonian},  include identical saturation coefficients $\beta$ for the two cavities, ensuring that the structural symmetry present in the linear eigenstates is preserved even in the nonlinear regime.

\begin{figure*}%
  \includegraphics[width=14cm]{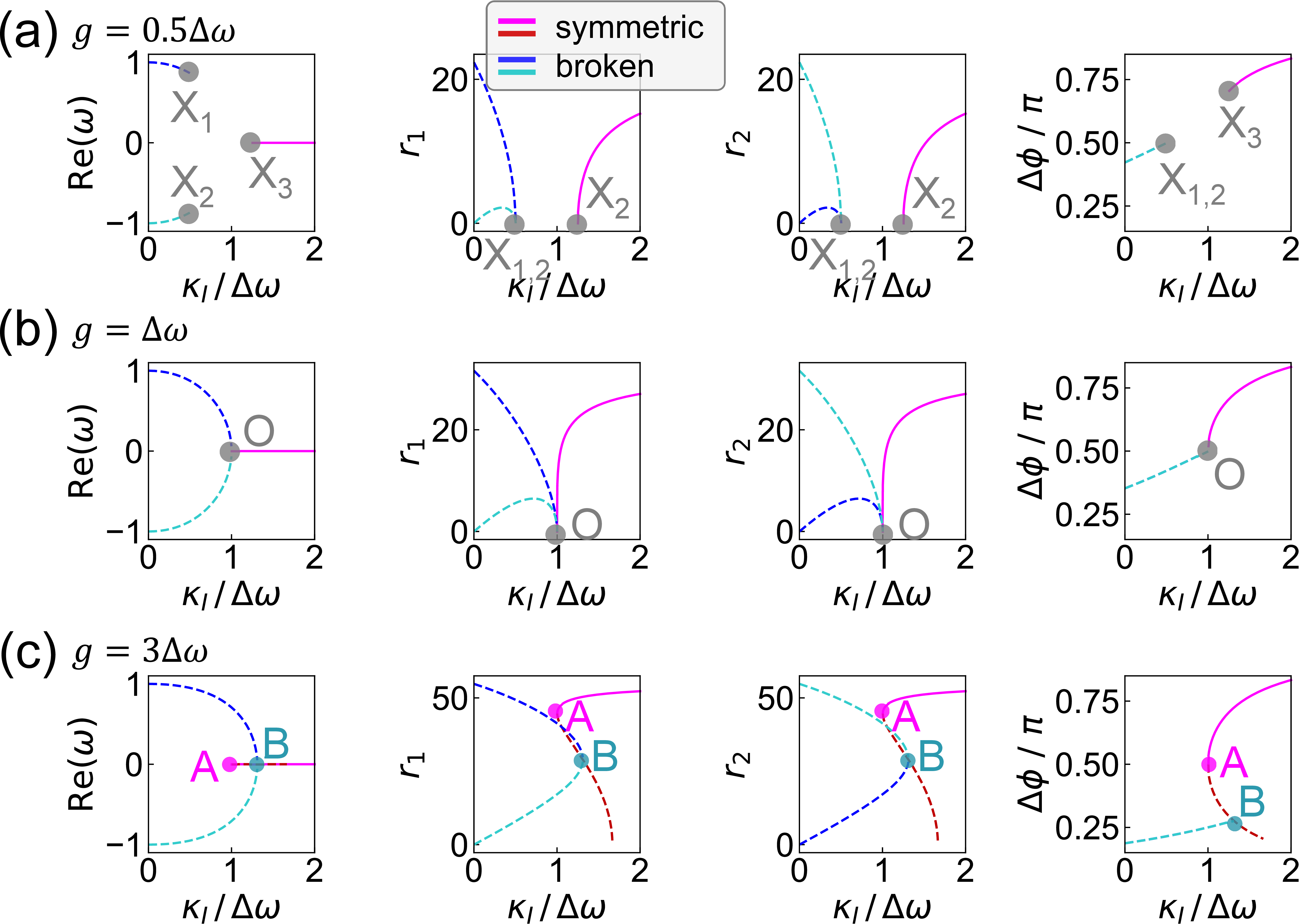}
  \caption{\label{fig_equilibrium}
    Frequency $\omega$, amplitudes $r_1=\sqrt{I_1}, r_2=\sqrt{I_2}$, and phase difference $\Delta \phi$ at the equilibrium points with nonlinear gain saturation $\beta = 10^{-3}$ and detuning $\Delta \omega = 1$ for three different gain values:
    (a) $g=0.5 \Delta \omega$, (b) $g=\Delta \omega$, and (c) $g=3 \Delta \omega$.
    The solid and dashed lines represent the stable and unstable equilibrium states, identified by the signs of the eigenvalues of the Jacobian matrix \eqref{eq:Jacobian} in Section \ref{sec:stability}.
  }
\end{figure*}

We now analyze the equilibrium states under the nonlinear regime ($\beta > 0$) with finite gain ($g > 0$), where the system exhibits steady-state lasing. 
The solid and dashed lines in Fig.~\ref{fig_equilibrium} indicate stable and unstable equilibrium states, respectively. A detailed discussion of the stability analysis is provided in Section~\ref{sec:stability}. The condition $I_1 > 0$ and $I_2 > 0$ implies lasing, and the onset of lasing corresponds to the point where $I_1=0$ or $I_2=0$ bifurcates from zero.

We first consider the regime $g = 0.5 \Delta \omega < \Delta \omega$, 
where the linear gain is insufficient to compensate for the energy dissipation in the regime near $\kappa_I = \Delta \omega$.
As shown in Fig.~\ref{fig_equilibrium}(a), lasing first occurs at $X_1, X_2, X_3$, which are far from the EP. This reflects the fact that dissipative coupling $\kappa_I$ governs interference between the spectral tails of the cavities, which in turn determines energy loss via Fano-like interference. In these regimes, destructive interference between the cavity modes suppresses energy leakage into the waveguide, effectively reducing the total loss and allowing lasing.

Figure~\ref{fig_equilibrium}(b) shows the results for $g = \Delta \omega$, at which the linear gain exactly compensates for the dissipative coupling loss at $\kappa_I = \Delta \omega$ ($\gamma = 0$). Here, $\kappa_I = \Delta \omega$ is the EP condition for the linear Hamiltonian in \eqref{eq:Hamiltonian_linear}. At this point, the intensities vanish ($I_1 = I_2 = 0$), indicating that the system is precisely at the lasing threshold. The third terms of \eqref{eq:nonlinear_phi1} and \eqref{eq:nonlinear_phi2} indicate that energy exchange between the two cavities is maximized when $\Delta \phi = \pm \pi/2$. At this point, the complex amplitudes $a_1$ and $a_2$ are injected into each other with a relative phase of $\pi/2$, meaning that their complex amplitudes are orthogonal in the complex plane. This interference does not facilitate energy recycling between the cavities, and thus fails to compensate for the dissipative loss. This situation can be interpreted as the inverse of Fano interference typically seen in electromagnetically induced transparency (EIT) \cite{Wang2020}. In Ref.~\cite{Wang2020}, an EP is formed between CW and CCW modes in a single microring cavity. Destructive interference between direct excitation and a recirculated pathway suppresses absorption, enabling transparent transmission. In contrast, our system involves destructive interference between modes of two coupled cavities. At the EP, the phase difference $\Delta \phi = \pm \pi/2$ leads to orthogonal coupling, preventing energy recycling, which maximizes the energy loss. This explains why $I_1 = I_2 = 0$ is satisfied exactly at point O. Since the field amplitudes are infinitesimal, nonlinear terms are negligible, and the system effectively remains in the linear regime. As a result, the equilibrium point O at $\kappa_I = \Delta \omega$ coincides with the EP of the linear Hamiltonian.
Importantly, the steady-state solutions around point O exhibit symmetric and broken branches, analogous to the eigenstates of the linear system. 
It is particularly surprising that the frequency $\omega$ and phase difference $\Delta \phi$ presented in Fig.~\ref{fig_equilibrium}(b) exhibit a striking resemblance to those derived from the linear analysis in Fig.~\ref{fig_eigenstates}.
Moreover, the phase difference becomes $\Delta \phi = \pm \pi/2$ at the EP, and the derivative $\partial \omega / \partial \kappa_I$ diverges, clearly manifesting the linear EP characteristics. The preservation of these properties in the presence of nonlinearity is surprising, indicating that the EP bifurcation pattern survives into the nonlinear regime when $g = \Delta \omega$.

Figure~\ref{fig_equilibrium}(c) presents the case $g = 3\Delta \omega$, where the nonlinear effect is significant. In the lasing state, the nonlinear bifurcation point O splits into two new bifurcation points, A and B. In the symmetric states presented as solid lines, the condition $I_1 = I_2$ always holds, ensuring that the symmetry of the intensity is preserved even above the lasing threshold. In contrast, for the broken states, the magnitude of the gain saturation effect varies between the two cavities, leading to a shift in the position of bifurcation point B along the $\kappa_I$ axis. Here, the solution with the minus sign in Eq. \eqref{eq:sol_symmetric} and both branches of the broken solution in Eq. \eqref{eq:sol_broken} coalesce exactly at point B ($\kappa_{I}=\sqrt{(\Delta \omega)^2 + \gamma^2 / 4}$). Remarkably, points A and B partly inherit some properties of the initial EP: Both points A and B exhibit $I_1 = I_2$, and the phase difference $\Delta \phi = \pm \pi/2$ holds only at point A. On the other hand, the divergence of $\partial \omega / \partial \kappa_I \to \infty$ occurs only at point B. Now, let us consider the characteristics of the ``nonlinear Hamiltonian'' in Eq.\eqref{eq:Hamiltonian} at the bifurcation points to discuss the connection between EPs and nonlinear bifurcation points. By substituting $\kappa_I=\Delta \omega$ and the equilibrium states at point A, $\boldsymbol{a}_\mathrm{A}(0) = \sqrt{\gamma / \beta}{[1, e^{-i\pi/2}]}^\top$, into \eqref{eq:Hamiltonian}, we find that the nonlinear Hamiltonian \eqref{eq:Hamiltonian} is defective at bifurcation point A:  
\begin{align}
 H(\boldsymbol{a}_\mathrm{A}(t)) \propto
  \begin{bmatrix}
  1 & -i \\
  -i & -1 \\
  \end{bmatrix}
\end{align}
On the other hand, we can see that $H(\boldsymbol{a}(t))$ is not defective at point B from $\kappa_{I}=\sqrt{(\Delta \omega)^2 + {(\gamma / 4)}^2}$ and $\boldsymbol{a}_\mathrm{B}(0) = \sqrt{\gamma / (2\beta)}{[1, e^{-i\theta_B}]}^\top$, where $\theta_B$ is the phase difference at point B:
\begin{align}
   H(\boldsymbol{a}_\mathrm{B}(t)) \propto 
   \begin{bmatrix}
   1 + i\frac{\gamma}{2\Delta \omega} & -i\sqrt{1 + \frac{\gamma^2}{4\Delta \omega^2} } \\
   -i\sqrt{1 + \frac{\gamma^2}{4\Delta \omega^2} } & -1 + i\frac{\gamma}{2\Delta \omega} \\
   \end{bmatrix}
\end{align}
This leads us to interpret point A as a nonlinear exceptional point. Remarkably, despite being in a high-intensity, strongly nonlinear regime, point A retains the essential properties of the linear EP: symmetry, phase chirality, and Hamiltonian defectiveness. This demonstrates that in the nonlinear regime, the EP bifurcates into two distinct bifurcation points, one of which—point A—should be recognized as a nonlinear EP.

\subsection{Stability analysis}
\label{sec:stability}

\begin{figure}%
  \includegraphics[width=5.cm]{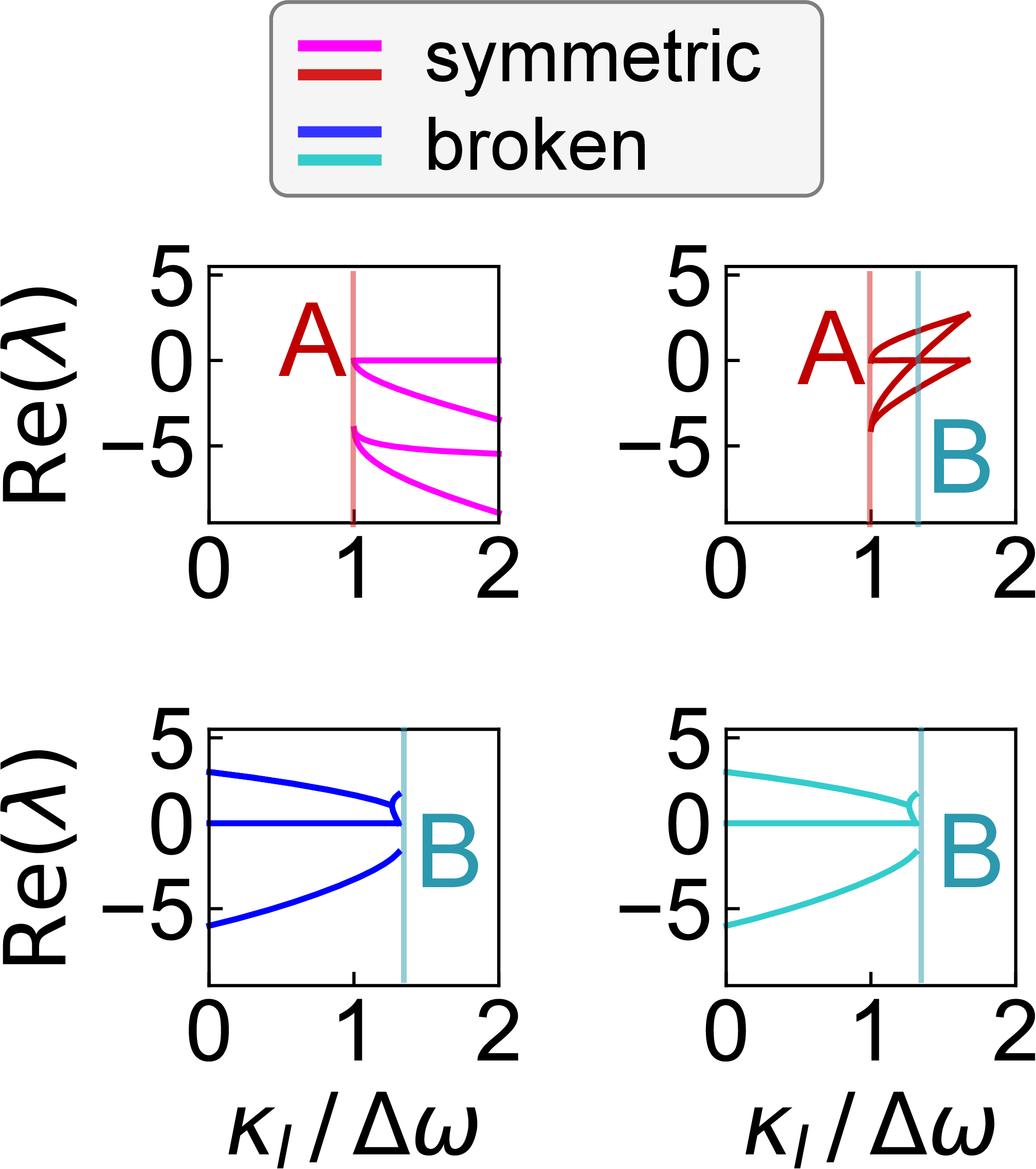}
  \caption{\label{linearStability}
		The eigenvalues of the Jacobian matrix for the equilibrium points at $g = 3 \Delta\omega$. The state in the upper left is stable, as all other eigenvalues have negative real parts, except for a single zero eigenvalue arising from global phase invariance.
    }
\end{figure}
To further clarify the physical relevance of the point A, we now perform a linear stability analysis of the steady-state solutions \cite{strogatz2018nonlinear}. 
In the following analysis, we consider the regime $g = 3\Delta\omega > \Delta\omega$, where the system is well above the lasing threshold.
The stability of each equilibrium point is determined by evaluating the eigenvalues of the Jacobian matrix $J$ \cite{Ji2023, Fischer2024}, which describes the local linearized dynamics around the equilibrium state $\boldsymbol{x}_0$. The Jacobian is defined as
\begin{align}
  \label{eq:Jacobian}
	J &:= {\left. \frac{\mathrm{d}}{\mathrm{d} \boldsymbol{x}} \boldsymbol{F}(\boldsymbol{x}) \right|}_{\boldsymbol{x}=\boldsymbol{x}_0}, \nonumber\\
	\boldsymbol{x} &= {\left[ r_1, r_2, \phi_1, \phi_2 \right]}^T, \quad \boldsymbol{F}(\boldsymbol{x}) = \frac{\mathrm{d}}{\mathrm{d} t} \boldsymbol{x}.
\end{align}
Substituting Eqs.~\eqref{eq:nonlinear_r1}–\eqref{eq:nonlinear_phi2} into this definition yields the Jacobian matrix:
%
\begin{align}
	J &=
	\begin{bmatrix}
	\gamma - 3 \beta r_1^2 & \kappa_I c & \kappa_I r_2 s & -\kappa_I r_2 s \\
	\kappa_I c & \gamma - 3 \beta r_2^2 & \kappa_I r_1 s & -\kappa_I r_1  s \\
	-\kappa_I \frac{r_2}{r_1^2} s & \kappa_I \frac{1}{r_1} s & -\kappa_I \frac{r_2}{r_1} c & \kappa_I \frac{r_2}{r_1} c \\
	-\kappa_I \frac{1}{r_2} s & \kappa_I \frac{r_1}{r_2^2} s & \kappa_I \frac{r_1}{r_2} c & -\kappa_I \frac{r_1}{r_2} c
	\end{bmatrix}, \nonumber\\
	c &= \cos\Delta \phi, \quad s = \sin\Delta \phi
\end{align}
%
The four panels in Fig.~\ref{linearStability} show the eigenvalues of the Jacobian matrix for $g = 3 \Delta \omega$, corresponding to the four equilibrium branches separated by points A and B in the middle panel of Fig.~\ref{fig_equilibrium}(c). These branches include two symmetric and two broken solutions on either side of the bifurcation points.
Since the original system is described by two complex amplitudes, the Jacobian matrix is real and $4 \times 4$. Each equilibrium point always has a zero eigenvalue due to the global phase invariance of the system. The stability is determined by the signs of the remaining eigenvalues.
The analysis reveals that only one of the symmetric equilibrium branches possesses eigenvalues with negative real parts (plotted as top left panel, the purple line in Fig.~\ref{linearStability}), indicating its stability. Crucially, point A is located at the termination of this stable branch and corresponds to a saddle-node bifurcation, where a pair of stable and unstable fixed points merge and annihilate. On the other hand, point B is associated with a pitchfork bifurcation where three unstable branches intersect. The bifurcation point A identified in our APT-symmetric system satisfies the definition of a nonlinear EP based on the nonlinear Hamiltonian defectiveness~\cite{Ji2023}, while also inheriting key characteristics of linear EPs—namely, eigenvalue coalescence, a $\pi/2$ phase difference, and Hamiltonian defectiveness.


\begin{figure}%
  \includegraphics[width=8cm]{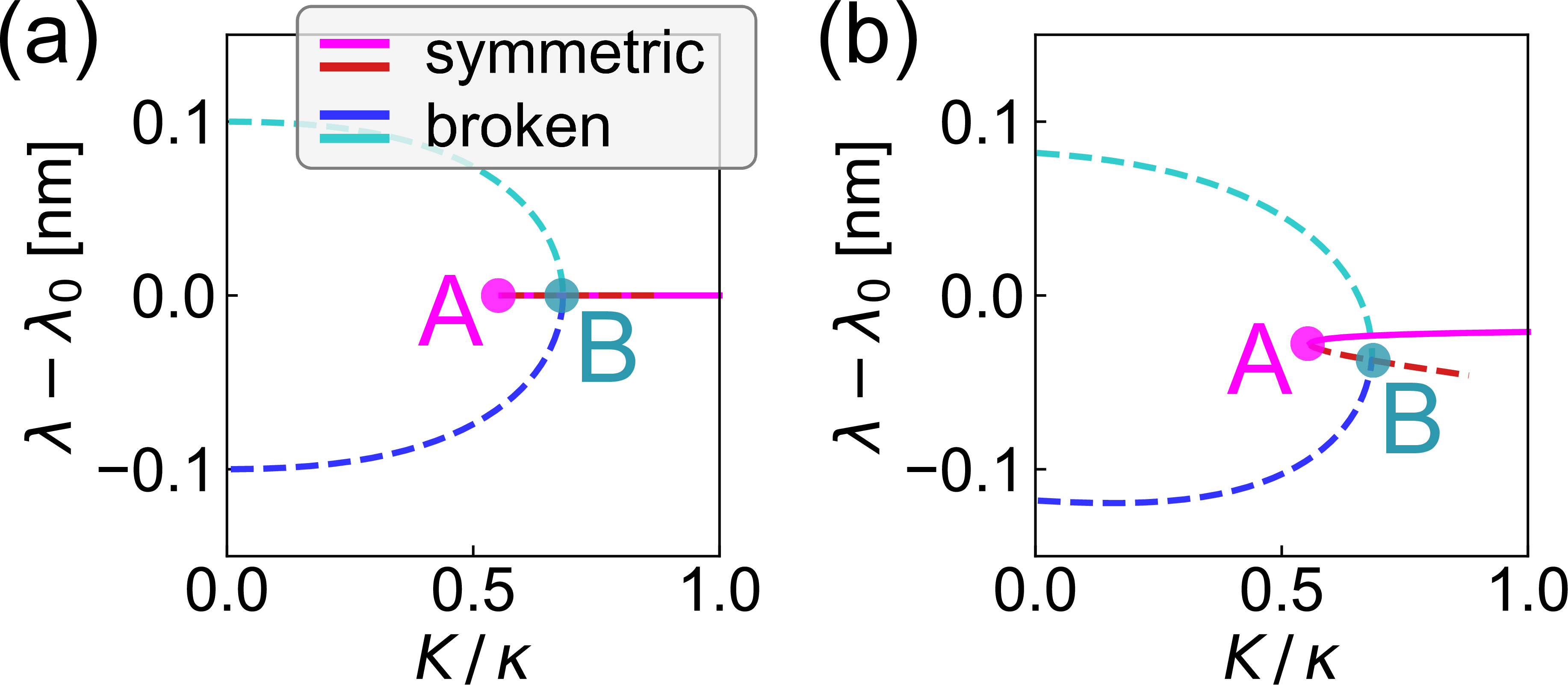}
  \caption{\label{fig_equiv_w_rateeq}
    The equilibrium wavelength of the coupled cavity system as a function of normalized coupling rate $K \, / \, \kappa$ with carrier dynamics in rate equation (Eqs. \eqref{eq:rate_eq_amp} and \eqref{eq:rate_eq_phase}) when the linewidth enhancement coefficient is set to (a) $\alpha = 0$ and (b) $\alpha = 0.1$. The other parameters are shown in Table~\ref{tab:rate_eq_params} in Supplemental Information~\ref{sec:rate_eq_params}. 
  }
\end{figure}

In practical semiconductor lasers, carrier dynamics significantly influence nonlinear effects and system stability. When carrier dynamics are taken into account, the dynamics of cavities 1 and 2 are described by the following set of rate equations:
\begin{align}
  \label{eq:rate_eq_amp}
  \frac{d a_{1,2}}{dt} &= \left[ i\omega_{1,2} - \kappa - K \right. \nonumber\\
  &\quad \left. + \frac{1 + i\alpha}{2} \beta \gamma_\parallel (n_{1,2} - n_0) \right] a_{1,2} + iK a_{2,1} \\
  \label{eq:rate_eq_phase}
  \frac{d n_{1,2}}{dt} &= P_{1,2} - \gamma_{\text{tot}} n_{1,2} - \beta \gamma_\parallel (n_{1,2} - n_0) |a_{1,2}|^2
\end{align}
Here, the terms related to white noise are neglected. In this model, the nonlinear effects of the carrier densities $n_{1,2}$ manifest as gain saturation. Moreover, the linewidth enhancement factor $\alpha$ contributes to a real-frequency shift that depends on the carrier density. The meaning of each parameter is summarized in Table~\ref{tab:rate_eq_params} in Supplemental Information~\ref{sec:rate_eq_params}. The total pump rate is set to three times the lasing threshold for a single cavity, and is consistent with the experimental conditions of the PT-symmetric laser system in Ref. \cite{Ji2023}, as detailed in Appendix \ref{sec:rate_eq_params}. This ensures that the system is in a condition analogous to the $g = 3\Delta \omega$ case analyzed with the Stuart-Landau model. The result shown in Fig.~\ref{fig_equiv_w_rateeq}(a) illustrates the equilibrium points under the condition $\alpha = 0$, which is consistent with a simple Stuart-Landau nonlinearity as shown in Fig.~\ref{fig_equilibrium}(c). Furthermore, the bifurcation point A located at the left edge of the symmetric-phase branch remains semi-stable even when $\alpha$ is nonzero. This indicates that the carrier dynamics do not affect the stability of the bifurcation point associated with the EP in APT symmetric systems. Here, the frequency detuning observed between the two symmetric phases originates from a difference in the steady-state carrier densities. 

The fact the linear EP in the APT-symmetric system transitions to a nonlinear bifurcation point that inherits its key properties—namely, eigenvalue coalescence, a $\pi/2$ phase difference, and Hamiltonian defectiveness, provides clear theoretical backing for the analogy that was implied in the previous work \cite{Takemura2021} between EPs and bifurcation points, as well as between linear eigenvalue dispersion and the average frequency. Furthermore, we reveal that the symmetries of the linear eigenstates (APT-symmetric and APT-broken) are also inherited in the nonlinear regime, a situation that differs from the previous findings in PT-symmetric systems. Furthermore, the EP-oriented bifurcation point in APT-symmetric systems is observable even under the influence of carrier dynamics in semiconductor lasers, and this is particularly important for semiconductor lasers (class-B lasers), which dominate most industrial applications from optical communications to laser processing and exhibit a wealth of complex nonlinear behaviors. This finding contrasts with the results in PT-symmetric systems (Ref. \cite{Ji2023}), in which relaxation oscillations near nonlinear bifurcation points arise from a mismatch in timescales between the total decay rate of the population inversion $\gamma_{\text{tot}}$ and the cavity decay rate $\kappa$. Specifically, asymmetries in pumping lead to carrier-induced detuning. Due to differences in relaxation times, this detuning is not fully suppressed, resulting in sustained oscillations. Conversely, in APT-symmetric systems, where the excitation strength is identical for both cavities, no steady oscillations originating from carrier-induced detuning occur. It should be noted, however, that discussing a physical correspondence in PT-symmetric systems is also possible. As pointed out in Ref. \cite{Ji2023}, it has been theoretically suggested that for class-A lasers, the EP can be stable, and in such a scenario, one could discuss the physical significance of EPs and eigenstates in the nonlinear regime (see Appendix \ref{sec:PT_w_StuartLandau}). Nevertheless, as we will see in the next section, the APT-symmetric case provides a more intimate connection to the nonlinear dynamics confirmed by time evolution.

\section{Time-domain dynamics}

\subsection{Time evolution of the system}

\begin{figure*}%
  \includegraphics[width=17cm]{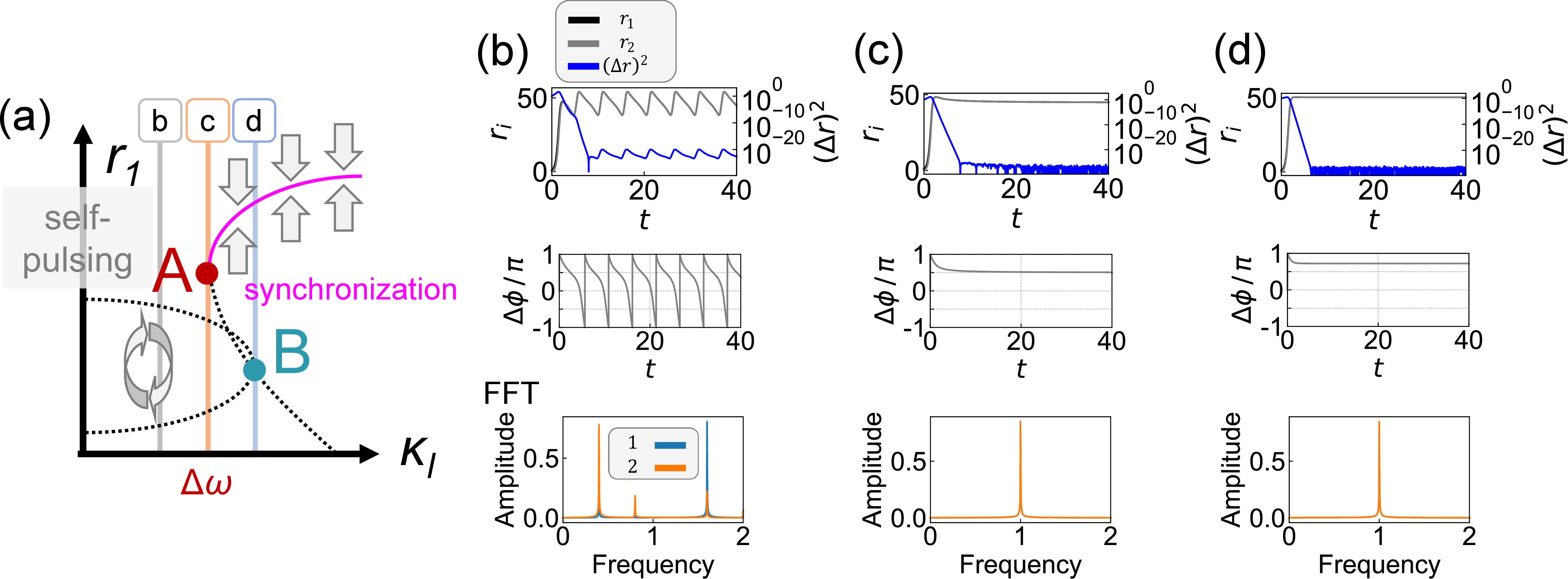}
  \caption{\label{fig_time_evolution}
    (a) Schematic diagram of the dynamics along the equilibrium points.  
    (b)-(d) Time evolution of the amplitudes and phases, and the Fourier spectrum for (b) $\kappa_I = 0.8 \Delta \omega$, (c) $\kappa_I = \Delta \omega$ (point A), and (d) $\kappa_I = 1.3096\Delta \omega$ (point B).    
  }
\end{figure*}

In the previous section, we investigated the correspondence between a linear APT symmetric system and a nonlinear open system by comparing the results of linear eigenvalue analysis and nonlinear steady-state analysis. However, the fixed points illustrated in Fig. \ref{fig_equilibrium} and the stability analysis in Fig. \ref{linearStability} are insufficient to fully characterize the actual dynamics of the system. In particular, the broken solutions are unstable and thus do not necessarily reflect the observable dynamical behavior in the time domain.
To verify the physical relevance of the equilibrium states and their stability properties, we now turn to the analysis of the system's time-domain dynamics. 
As in Section~\ref{sec:stability}, we consider the case $g = 3\Delta\omega$ as shown in Fig.~\ref{fig_equilibrium}(c), where all the modes are assumed to be lasing. By numerically solving the full nonlinear equations of motion, we can observe how the system evolves from a given initial condition and determine whether it indeed converges to the theoretically predicted equilibrium states. The dynamics implied by the stability analysis of the system's equilibrium points are depicted in Fig. \ref{fig_time_evolution}(a). Here, the solid and dashed lines indicate the stable and unstable points, respectively. In the regime where $\kappa_I < \Delta \omega$, no stable equilibrium points exist. On the other hand, for $\kappa_I > \Delta \omega$, only the symmetric state remains stable, meaning that only this state should be lasing. At bifurcation point A, one stable and one unstable equilibrium point merge and disappear.

Figures \ref{fig_time_evolution}(b) to (d) show the time evolution of the amplitudes and phases calculated using the Runge-Kutta RK4 method with $\Delta t = 0.01, 0<t<T=500$, as well as the Fourier spectrum of the system with the initial condition $\boldsymbol{a}(t=0) = {[1, 0]}^\top$. The Fourier spectrum is obtained by applying the fast Fourier transform (FFT) from the time evolution data with $T/2 < t < T$.
For $\kappa_I < \Delta \omega$, the field cannot remain at the equilibrium point, leading to mode beating in the time evolution. This results in the appearance of the sub-peak (the center peak between three peaks) in the Fourier spectrum, as shown in Fig. \ref{fig_time_evolution}(b). Remarkably, the amplitude difference between the two cavities, $|r_1 - r_2|$, vanishes with time, showing a behavior different from the unstable equilibrium state with unequal amplitudes described in \eqref{eq:sol_broken}. 
In contrast, for $\kappa_I > \Delta \omega$, only a single-frequency oscillation occurs with $r_1 = r_2$, as confirmed in Fig. \ref{fig_time_evolution}(d). The amplitude and phase difference in the time evolution are consistent with the ``symmetric states'' in Fig. \ref{fig_equilibrium}(c). When $\kappa_I = \Delta \omega$, the equilibrium states satisfy $r_1 = r_2$ and $\Delta \phi = \pi /2$, confirming that the mode at point A is excited as intended (Fig. \ref{fig_time_evolution}(c)). 
Consequently, our numerical results are consistent with the expected behavior from the stability analysis shown in Fig. \ref{fig_time_evolution}(a) and indicate that bifurcation point A, which inherits the properties of EPs, can be identified as the phase transition point between the self-pulsing oscillation and synchronization.

\subsection{Convergence of amplitude differences}

\begin{figure}%
  \includegraphics[width=8cm]{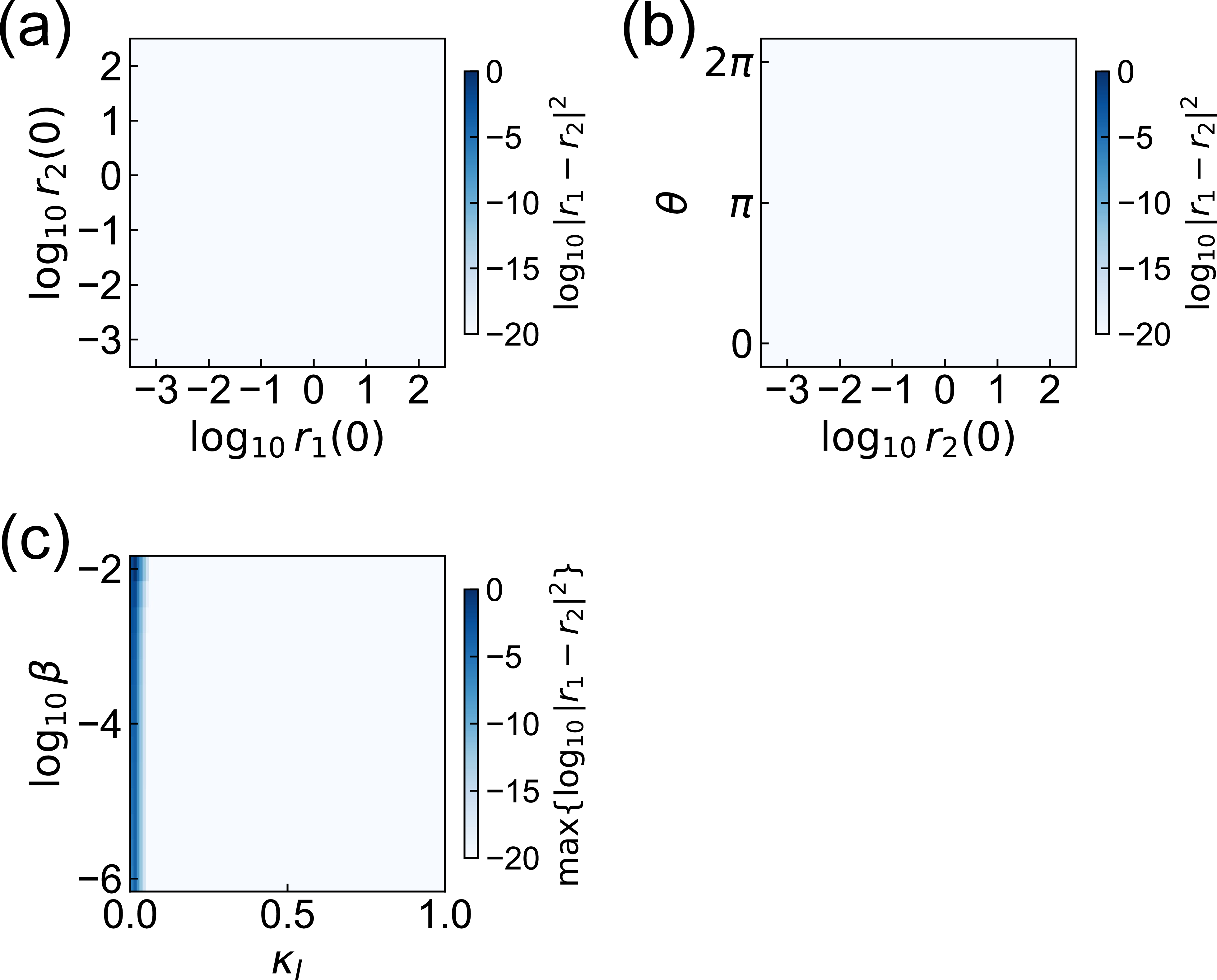}
  \caption{\label{fig:sweep_params}
    The value of $\log_{10}|r_1 - r_2|^2$ at the end time of numerical integration $T_\mathrm{end}$ for each parameter setting.
    (a) The case where $r_1(0)$ and $r_2(0)$ are varied according to Table \ref{tab:table1} under the condition $\Delta \phi (0) = 1$.
    (b) The case where $r_2(0)$ and $\Delta \phi $ are varied under the condition $r_1 = 1$. In both cases (a) and (b), the calculations were performed with $\kappa_I = 0.1$ and $\beta = 1\times10^{-3}$.
    (c) The case where all parameters $r_1(0)$, $r_2(0)$, $\Delta \phi(0)$, $\kappa_I$, and $\beta$ are varied. For each $(\kappa_I, \beta)$, the maximum value within the corresponding group of $(r_1(0), r_2(0), \Delta \phi(0))$ is plotted.
    The Runge-Kutta method was used with a timestep of $\Delta t = 0.01$ and an end time of $T_\mathrm{end} = 1000$.
  }
\end{figure}

\begin{table}[b]
\caption{\label{tab:table1}%
Summary of parameter ranges and discretization.
}
\begin{ruledtabular}
\begin{tabular}{lcrr}
\colrule
Variable & Range & Grid points & Scale \\
\hline
$r_1(0)$ & $10^{-3}$ to $10^{2}$ & 6 & logarithmic \\
$r_2(0)$ & $10^{-3}$ to $10^{2}$ & 6 & logarithmic \\
$\Delta \phi(0)$ & $0$ to $2\pi$ & 21 & linear \\
$\kappa_I$ & $0$ to $1$ & 101 & linear \\
$\beta$ & $10^{-6}$ to $10^{-2}$ & 13 & logarithmic
\end{tabular}
\end{ruledtabular}
\end{table}

Despite the asymmetric amplitude profile of the broken solution (Eq.~\eqref{eq:sol_broken}), the system dynamics shown in Fig.~\ref{fig_time_evolution} (b) suggest that the amplitudes still converge to $r_1 = r_2$ even in the regime $\kappa_I < \Delta\omega$. This observation implies that the long-time dynamics of the system effectively collapse onto a two-dimensional phase space defined by $(r, \Delta\phi)$, irrespective of the value of $\kappa_I$. If this reduction indeed holds, it has significant implications: it supports the interpretation that the beat oscillations observed in Fig.~\ref{fig_time_evolution} (b) constitute a limit cycle, and it establishes a direct dynamical correspondence between the nonlinear system and its linear APT counterpart.

While a rigorous analytical proof of the convergence to $r_1 = r_2$ remains challenging (a detailed numerical study is provided in Supplemental \ref{sec:appendix_convergence}), we can numerically confirm the convergence of $|r_1 - r_2|^2$ over a wide range of initial conditions and parameters. Here, we consider the case $g = 3\Delta\omega$, where all modes are assumed to be lasing.
We systematically varied the initial conditions $\boldsymbol{a}(t=0) = {\left[r_1(0), r_2(0)e^{-i\Delta \phi(0)}\right]}^\top$ and system parameters $(\kappa_I, \beta)$ based on the settings listed in Table~\ref{tab:table1}. We computed the amplitude difference $|r_1(T_\mathrm{end})-r_2(T_\mathrm{end})|$ at the final integration time $T_\mathrm{end}$. Table~\ref{tab:table1} summarizes the range, number of grid points, and grid scaling method for each parameter. Specifically, logarithmic spacing was applied for $r_1(0)$, $r_2(0)$, and $\beta$, whereas linear spacing was employed for $\Delta \phi(0)$ and $\kappa_I$. Note that results for cases with $r_1(0), r_2(0) > 100$ or $\beta > 10^{-2}$ were omitted, as numerical errors associated with the timestep size $\Delta t$ became significant and affected computational feasibility.

The logarithm of $|r_1(T_\mathrm{end})-r_2(T_\mathrm{end})|^2$ evaluated at $T_\mathrm{end}$ is shown in Fig.\ref{fig:sweep_params}. The numerical integration was performed using a fourth-order Runge-Kutta method with timestep $\Delta t = 0.01$ and final time $T_\mathrm{end}=1000$. Figure \ref{fig:sweep_params}(a) shows the results obtained by sweeping $r_1(0)$ and $r_2(0)$ while fixing $\Delta \phi(0)=1$, whereas Fig.~\ref{fig:sweep_params}(b) presents the results from sweeping $r_2(0)$ and $\Delta \phi(0)$ with $r_1(0)=1$. In both cases, the parameters were set to $\kappa_I = 0.1$ and $\beta = 10^{-3}$. Since $|r_1 - r_2|^2$ remains below $10^{-20}$ for most parameter sets, we conclude that the system has effectively reached convergence.
Figure \ref{fig:sweep_params}(c) summarizes the results when all five parameters $(r_1(0), r_2(0), \Delta \phi(0), \kappa_I, \beta)$ are simultaneously swept. For each pair of $(\kappa_I, \beta)$ values, we plot the maximum value of $|r_1(T_\mathrm{end})-r_2(T_\mathrm{end})|$ observed among the corresponding initial condition group. Overall, the system exhibits convergence across a broad parameter space. In the regime of weak coupling $\kappa_I < 0.05$, $|r_1(T_\mathrm{end})-r_2(T_\mathrm{end})|$ remains finite, which is attributed to insufficient integration time for complete convergence under such weak coupling conditions. Consequently, even within the self-pulsing regime, the system robustly exhibits convergence toward $|r_1(t) - r_2(t)| \to 0$ as $t \to \infty$ over a wide range of initial conditions and parameters.
This result demonstrates that the dynamics globally converge to the state with equal amplitudes, even when starting from initial conditions corresponding to the unstable `broken' solutions, which are generally characterized by unequal amplitudes ($r_1 \neq r_2$) as shown in Eq.~\eqref{eq:sol_broken}.

\subsection{Limit cycle oscillations and infinite-period bifurcation}
\label{sec:limit_cycle}

\begin{figure*}%
  \includegraphics[width=15cm]{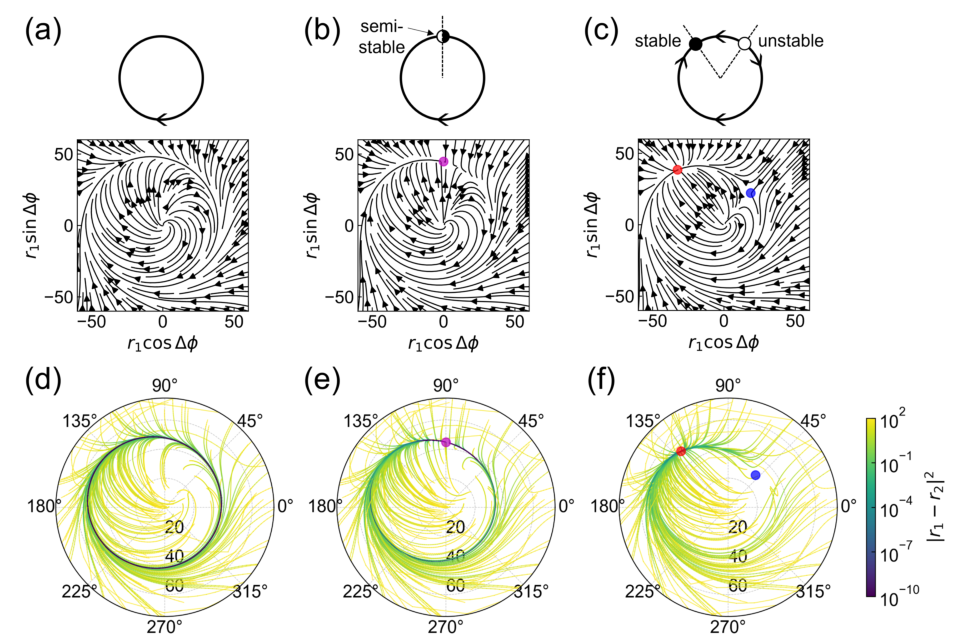}
  \caption{\label{fig:limitcycle}
    (a)-(c) Vector fields of the system in the $(r_1, \Delta\phi)$ plane for (a) $\kappa_I=0.8\Delta\omega$, (b) $\kappa_I=\Delta\omega$ and (c) $\kappa_I=1.3096\Delta\omega$, corresponding to the same parameter sets as in Figs. 5(a), (b) and (c). The top of each panel shows a schematic illustration of the dynamical behavior.
    The magenta dot on (b) indicates the bifurcation point A shown in Fig. \ref{fig_equilibrium}(c), where the system converges to a stable fixed point, where the system converges to a semi-stable fixed point. The red and blue dot on Fig. \ref{fig_equilibrium}(c) indicate the stable and unstable fixed points of the symmetric state, and the latter corresponds to the bifurcation point B in Fig. \ref{fig_equilibrium}(c). (d)-(f) Time evolution trajectories of the system in the $(r_1, \Delta\phi)$ plane for (d) $\kappa_I=0.8\Delta\omega$, (e) $\kappa_I=\Delta\omega$ and (f) $\kappa_I=1.3096\Delta\omega$.
    The polar plots represent $r_1$ as the radial coordinate and $\Delta\phi = \phi_2 - \phi_1$ as the angular coordinate. The color bar indicates the squared amplitude difference $|r_1 - r_2|^2$.
  }
\end{figure*}

Since the system converges to the dynamics with $r_1 = r_2$, the steady-state behavior can be described using two variables, $r = r_1 = r_2$ and $\Delta\phi$. 
Under the assumption $r_1 = r_2$, the equations of motion in polar coordinates, Eqs.~\eqref{eq:nonlinear_r1}-\eqref{eq:nonlinear_phi2}, can be significantly simplified, as discussed in the following:
\begin{align}
  \label{eq:nonlinear_r}
  \frac{\mathrm{d}}{\mathrm{d}t} r &= (\gamma-\beta r^{2}) r + \kappa_I r\cos\Delta\phi,\\
  \label{eq:nonlinear_dphi}
  \frac{\mathrm{d}}{\mathrm{d}t} \Delta\phi&=-2(\Delta\omega-\kappa_I\sin\Delta\phi)
\end{align}
We consider the case where \(\Delta\omega > \kappa_I\). Since this is a planar system defined over a cylinder \(r \geq 0\), \(\Delta\phi \in \mathbf{R} \mod 2\pi\), the Poincaré-Bendixson theorem \cite{strogatz2018nonlinear} can be applied to establish the existence of periodic orbits. We first define a closed, positively invariant region in the phase space. The radial equation satisfies \(\dot{r} \leq r(\gamma - \beta r^2 + \kappa_I)\), and the right-handed term becomes negative when \(r > R_{\max} := \sqrt{(\gamma + \kappa_I)/\beta}\). Thus, the trajectory cannot escape beyond \(r = R_{\max}\). The boundary at \(r = 0\) is invariant due to the multiplicative factor of \(r\), ensuring trajectories remain in the region \(0 \leq r \leq R_{\max}\). Hence, this annular region forms a compact, positively invariant set.

We next examine the existence of fixed points within this region. Setting the right-hand sides on Eqs.~\eqref{eq:nonlinear_r} and \eqref{eq:nonlinear_dphi} equal to zero yields the following equations:
\begin{align}
  \gamma - \beta r^2 + \kappa_I \cos \Delta \phi = 0, \quad \Delta\omega = \kappa_I \sin \Delta \phi
\end{align}
The second equation implies \(\sin\Delta\phi = \Delta\omega / \kappa_I\), which has no real solution under the assumption \(\Delta\omega > \kappa_I\). Therefore, no equilibrium point exists in the phase space. With a compact invariant region and no fixed points, the Poincaré-Bendixson theorem guarantees that any trajectory starting in this region must asymptotically approach a closed orbit. This establishes the existence of a limit cycle. The rotational equation for \(\Delta\phi\) is strictly monotonic, implying that the phase difference circulates continuously, and thus the periodic orbit corresponds to equal amplitude oscillations with non-trivial phase dynamics. Since the radial dynamics include a cubic damping term, the limit cycle is attracting. 

Figure \ref{fig:limitcycle} (a)-(c) show the vector fields of the system in the ($r_1, \Delta\phi$) plane for three distinct regimes, calculated with $\Delta t = 0.01$ and $0 < t < 30$. For $\kappa_I < \Delta\omega$ (a), the vector field indicates the presence of a stable limit cycle. Conversely, for $\kappa_I > \Delta\omega$ (c), the limit cycle is broken, and all trajectories converge to a stable fixed point (indicated by the red dot), which corresponds to the symmetric state with $r_1 = r_2$. The nearby unstable fixed point (blue dot) corresponds to the bifurcation point B. At the critical condition $\kappa_I = \Delta\omega$ (b), the system evolves toward the semi-stable fixed point A. Although previously identified as a saddle-node bifurcation, a more precise interpretation is that it corresponds to an infinite-period bifurcation where the limit cycle is annihilated. The term "semi-stable" is used to describe the dynamics at point A, as trajectories may either slowly approach or move away from it over an infinite timescale.

The time evolution trajectories illustrated in Figs. \ref{fig:limitcycle} (d)-(f) confirm the dynamical behavior predicted by the vector fields. For these plots, 200 initial values were sampled from uniform distributions: $r_1(0), r_2(0) \in \mathcal{U}(0, 100)$ and $\phi_1(0), \phi_2(0) \in \mathcal{U}(0, 2\pi)$. These polar plots represent $r_1$ as the radial coordinate and $\Delta\phi$ as the angular coordinate, with the color bar indicating the squared amplitude difference $|r_1 - r_2|^2$. For all initial conditions sampled, as the trajectory evolves, the amplitude difference between $r_1$ and $r_2$ diminishes, indicating that the system dynamics asymptotically reduce to the two-dimensional form described by Eqs.~\eqref{eq:nonlinear_r} and \eqref{eq:nonlinear_dphi}. These results visually demonstrate that, in the presence of gain saturation nonlinearity, the APT-symmetric system exhibits both a limit-cycle phase ($\kappa_I < \Delta\omega$) and a synchronized state ($\kappa_I > \Delta\omega$), with the bifurcation point A clearly serving as the boundary between the two regimes.

\subsection{Beat oscillation at weak coupling regime}

\begin{figure}%
  \includegraphics[width=5cm]{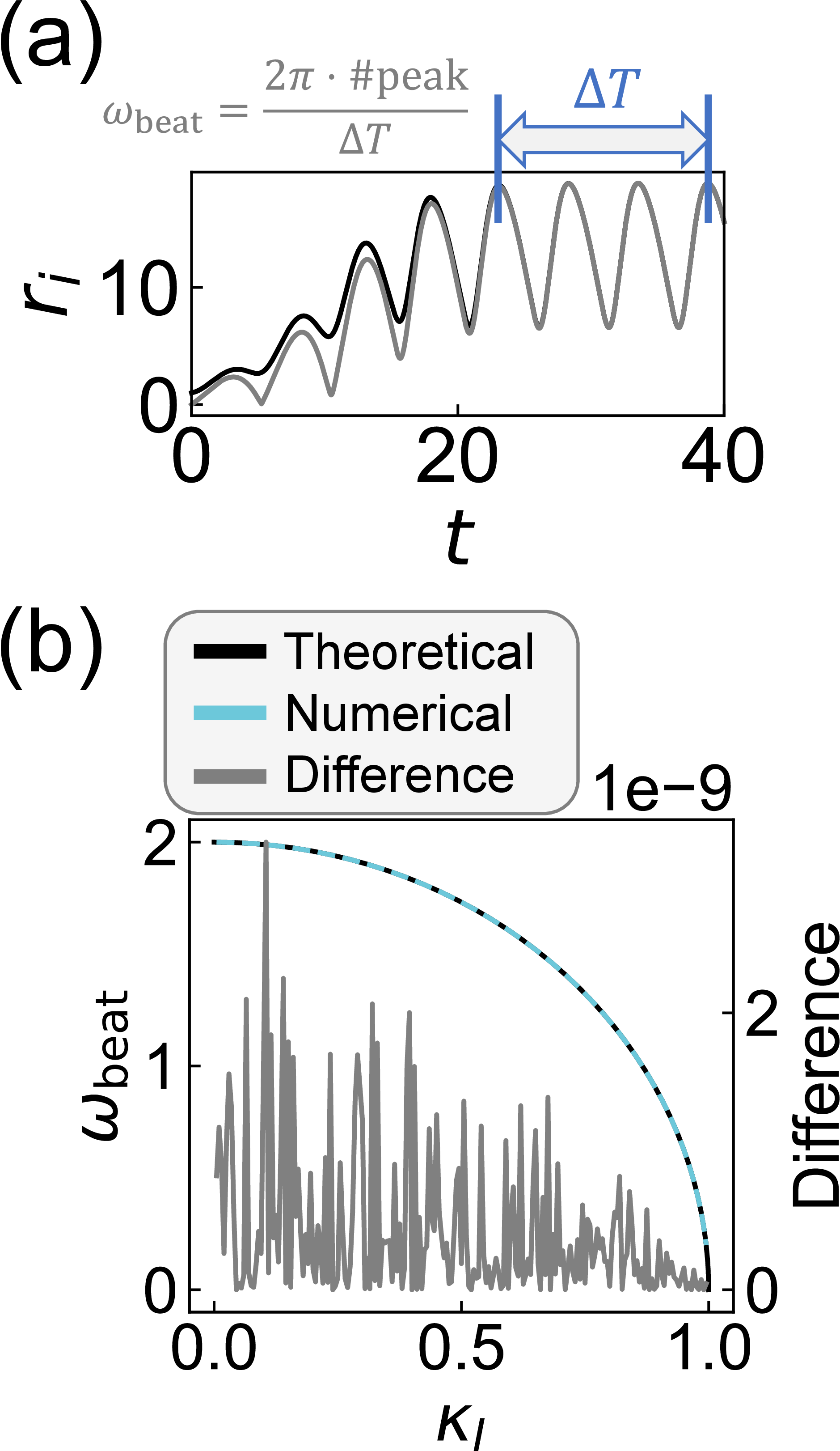}
  \caption{\label{fig:beat_frequency}
  (a) The way to calculate the beat frequency from the time evolution data. We count the number of peaks in the amplitude $r$ in the range $500 < t < 1000$ and divide it by the time interval between the first and last peaks.
  (b) Beat frequency \( \omega_\mathrm{beat} = 2\pi / T_\mathrm{beat} \) as a function of \( \kappa_I \) for $g=\Delta \omega$. 
  }
\end{figure}

Given that the system exhibits a limit cycle as established in the previous section, and that the dynamics on this cycle satisfy the condition $r_1=r_2$ for $t\to\infty$, we can analytically derive the associated beat frequency. The period of the beat oscillation, $T_\mathrm{beat}$, and the beat frequency $\omega_\mathrm{beat} = 2\pi / T_\mathrm{beat}$ can be derived from the phase difference dynamics in Eq.~\eqref{eq:nonlinear_dphi}.
\begin{align}
	\label{eq:nonlinear_one_cycle}
	T_\mathrm{beat} &= \int_0^{2\pi} \frac{\mathrm{d} t}{|\mathrm{d} \Delta \phi|} |\mathrm{d} \Delta \phi| \nonumber \\
	&= \frac{1}{2} \int_0^{2\pi} \frac{\mathrm{d} \phi}{\Delta \omega - \kappa_I \sin \phi} \nonumber \\
	&= \frac{\pi}{ \sqrt{(\Delta \omega)^2 - \kappa_I^2} } \\
  \label{eq:beat_frequency}
  \omega_\mathrm{beat} &= \frac{2\pi}{T_\mathrm{beat}} = 2\sqrt{(\Delta \omega)^2 - \kappa_I^2}.
\end{align}
The beat frequency $\omega_\mathrm{beat}$ is consistent with the difference of oscillation frequencies on two Kuramoto oscillators, whose oscillation frequencies are given by \cite{10.1143/PTP.79.1069}
\begin{align}
  \label{eq:avefreq1} 
  \langle \dot{\phi_{1}} \rangle &= \omega_0 + \sqrt{ {(\Delta \omega)}^2 - \kappa_I^2 }, \\
  \label{eq:avefreq2} 
  \langle \dot{\phi_{2}} \rangle &= \omega_0 - \sqrt{ {(\Delta \omega)}^2 - \kappa_I^2 }. 
\end{align}
This correspondence arises because, in the phase evolution equations Eq. \eqref{eq:nonlinear_phi1} and \eqref{eq:nonlinear_phi2}, setting $r_1 = r_2$ directly reduces the two coupled Kuramoto oscillators:
\begin{align}
  \label{eq:phase_reduction1} \frac{\mathrm{d}}{\mathrm{d} t} \phi_1 &= -\omega_1 + \kappa_I \sin(\phi_2 - \phi_1), \\
  \label{eq:phase_reduction2} \frac{\mathrm{d}}{\mathrm{d} t} \phi_2 &= -\omega_2 + \kappa_I \sin(\phi_1 - \phi_2)
\end{align}
Figure \ref{fig:beat_frequency} shows the beat frequency \( \omega_\mathrm{beat} = 2\pi / T_\mathrm{beat} \), calculated from the analytical expression \eqref{eq:nonlinear_one_cycle} and the numerical simulations, as a function of \( \kappa_I \). Numerical simulations are performed for \( \kappa_I \) values ranging from 0.005 to 0.995 in increments of 0.005, and the beat frequency \( \omega_\mathrm{beat} = 2\pi / T_\mathrm{beat} \) is computed using data obtained via the Runge-Kutta RK4 method with \( \Delta t = 0.01 \) and \( t_\mathrm{max} = 1000 \) and the following equation:
\begin{align}
	\label{eq:beatfreq_numerical}
	\omega_\mathrm{beat} = \frac{2\pi \cdot \# \mathrm{peak}}{\Delta T}
\end{align}
where \( \# \mathrm{peak} \) is the number of peaks in the amplitude \( r \) in the range \( 500 < t < 1000 \), and \( \Delta T \) is the time interval between the first and last peaks.
The beat frequency was calculated from the data in the range \( 500 < t < 1000 \).
The values obtained from the analytical expression and numerical simulations show excellent agreement, as illustrated in Fig.~\ref{fig:beat_frequency}(b). The difference between the two is confirmed to be on the order of $10^{-9}$ or less, which is negligibly small. This strong agreement demonstrates that the beat frequency derived under the condition $r_1 = r_2$ accurately explains the real dynamics of the system.

It is known that in linear APT-symmetric systems without gain saturation, beat oscillations arise as a manifestation of coherent energy exchange between the two modes \cite{Choi2018}. These beatings are a direct consequence of the APT symmetry, which allows for the exchange of energy while conserving the energy difference. Interestingly, the limit cycle oscillations discussed in the previous section apparently preserve this coherent energy exchange mechanism, even in the presence of gain saturation nonlinearity. 
In the following, we demonstrate the correspondence between the linear beat dynamics in APT systems and the nonlinear limit cycles that emerge in the presence of gain saturation. In a linear APT system, the eigenfrequencies split into a pair of conjugate modes owing to APT symmetry, and the frequency difference between these modes leads to observable beating oscillations. Denoting the eigenfrequencies derived from Eq.~\eqref{eq:eigenvalue} as $\omega_+$ and $\omega_-$, the time evolution of the field amplitude can be written as a superposition of the two eigenmodes:
\begin{align}
  a(t) &= A_+ e^{-i\omega_+ t} + A_- e^{-i\omega_- t}\\
  \therefore
  |a(t)|^2 &= |A_+|^2 + |A_-|^2 \nonumber\\
  &\quad + 2\mathrm{Re}(A_+ A_-^* e^{-i(\omega_+ - \omega_-)t})
\end{align}
This expression reveals that the beating frequency $\omega_{\mathrm{beat}}^{\mathrm{APT}}$ is determined by the real part of the frequency splitting:
\begin{align}
  \omega_\mathrm{beat}^{\mathrm{APT}} &= |\omega_+ - \omega_- |
  = 2\sqrt{ (\Delta \omega)^2 - \kappa_I^2 }
\end{align}
When a weak gain saturation nonlinearity is introduced into a linear APT-symmetric system, the amplitude difference $|r_1 - r_2|$ rapidly vanishes as $t \to \infty$, and the system dynamics transitions into a limit cycle. Remarkably, the beating frequency $\omega_\mathrm{beat}$ in Eq.~\eqref{eq:beat_frequency} remains identical to that of the original linear APT system, i.e., $\omega_\mathrm{beat} = \omega_\mathrm{beat}^{\mathrm{APT}}$. 
The difference is that while the linear beating appears as a pure sinusoidal oscillation, the nonlinear dynamics exhibit distorted trajectories, as illustrated in Fig.~\ref{fig:limitcycle} and Eqs.~\eqref{eq:nonlinear_r} and \eqref{eq:nonlinear_dphi}.
The beating oscillations observed in both linear and nonlinear APT systems inherit the underlying energy exchange mechanism dictated by APT symmetry, despite the presence of nonlinearity. Unlike relaxation oscillations in injection-locked lasers \cite{siegman1986lasers}, which is induced by carrier-photon population dynamics, these APT-induced beats arise from coherent interference effects governed by symmetry.

\subsection{Fourier spectrum of the time evolution}

\begin{figure}%
  \includegraphics[width=7.6cm]{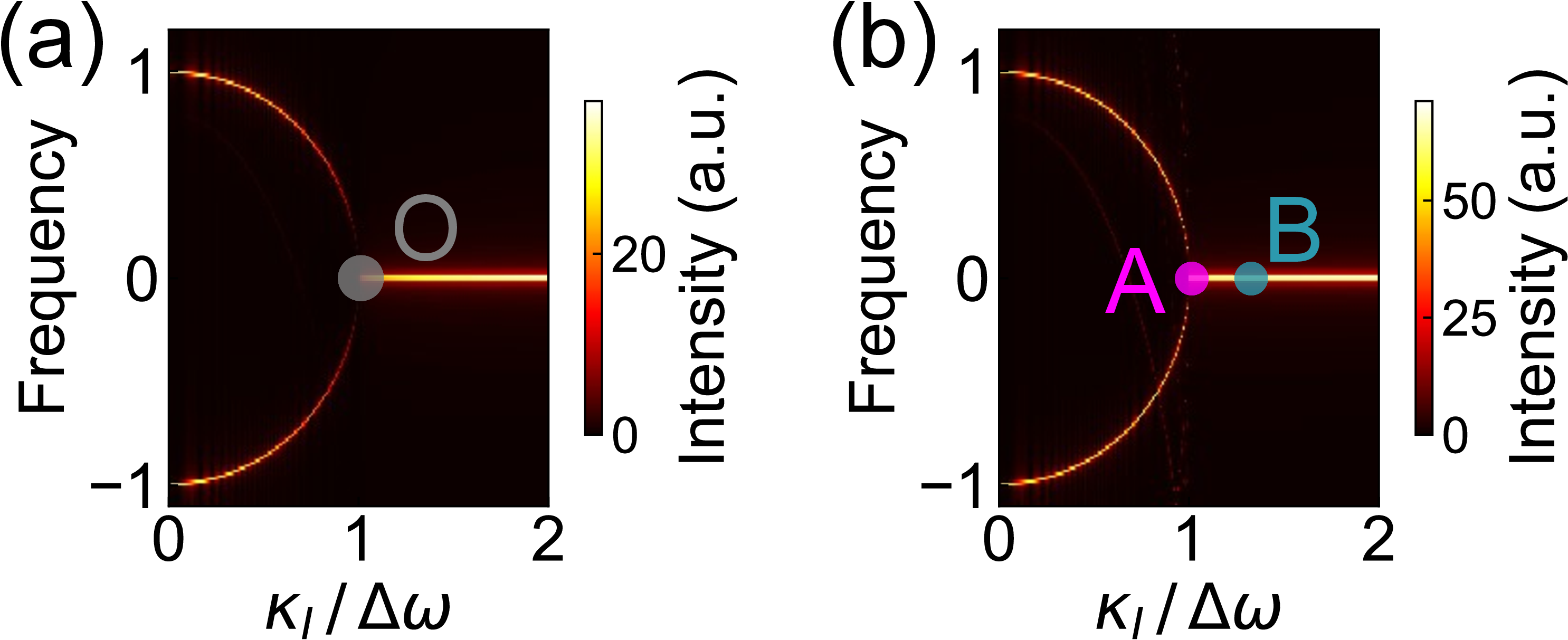}
  \caption{\label{fig_fft}
    The Fourier spectrum of the time evolution as a function of the coupling strength $\kappa_I$ for (a) $g=\Delta \omega$ and (b) $g=3\Delta \omega$.
  }
\end{figure}
Fourier spectrum of the time evolution as a function of the coupling strength can be depicted as the heatmap in Fig. \ref{fig_fft} for $\kappa_I$ for (a) $g=\Delta \omega$ and (b) $g=3\Delta \omega$. 
Due to the convergence of $|r_1 - r_2|$, the frequency shift of the broken state shown in Fig. \ref{fig_equilibrium}(c) is not confirmed in the Fourier spectrum. Instead, the oscillation is dominated by the beat frequencies corresponding to those of the Kuramoto oscillators, as given by Eqs. \eqref{eq:avefreq1} and \eqref{eq:avefreq2}.
Remarkably, the expressions for the oscillation frequencies in Eqs.\eqref{eq:avefreq1} and \eqref{eq:avefreq2} are mathematically identical in form to the eigenvalues of the linear Hamiltonian in Eq.\eqref{eq:eigenvalue}. Consequently, even above the lasing threshold, the peak frequencies observed in the Fourier spectrum coincide with the eigenfrequencies of the linear Hamiltonian given by Eq.\eqref{eq:eigenvalue}. Moreover, point A acts as a branching point in the spectrum, similar to the EP in the linear system.

\section{Discussion}

This study clarified a fundamental connection between EPs in linear APT-symmetric systems and bifurcation points in nonlinear dissipatively coupled cavities. Our findings reveal that the bifurcation point at the lasing threshold retains key features of the linear EP, such as amplitude symmetry, a $\pi/2$ phase difference between cavities, and a defective Hamiltonian structure, despite the presence of nonlinear gain saturation effects.
Importantly, our analysis demonstrated that in the nonlinear regime above threshold, the EP bifurcates into two distinct points (labeled A and B), among which point A inherits core EP characteristics.
Moreover, the bifurcation point A is physically accessible and governs a dynamical phase transition from a limit-cycle oscillation to synchronization, which is a consequence of the properties of the APT-symmetric system.

The phase transition from limit-cycle to synchronization is a common phenomenon in injection-locked laser systems, where synchronization occurs via a nonlinear transition as a function of injection intensity.
In conventional injection-locking studies \cite{siegman1986lasers, 1071166}, nonlinear dynamics such as synchronization, bifurcation, and chaos have been extensively discussed. However, the precise characterization of synchronization termination points, especially their connection to linear EPs, has remained largely unexplored. Our results establish that the synchronization termination point in injection-locked lasers can be identified with a bifurcation point that inherits EP properties — an insight not explicitly recognized in previous literature. Specifically, in our APT-symmetric system, the lasing threshold coincides with a linear EP where the eigenvalues and eigenvectors coalesce, accompanied by a phase difference of $\pi/2$. After surpassing the threshold, synchronization emerges as the system stabilizes onto a symmetric branch, while self-pulsing dynamics occur below the threshold.  A similar picture can be drawn for injection-locked lasers: near the locking threshold, the phase difference between the master and slave lasers becomes fixed, and the system transitions from a beat oscillation to full locking. Our work suggests that the bifurcation points observed in injection-locked lasers — typically treated purely within nonlinear dynamical systems theory — can be understood more fundamentally as the EP above the lasing threshold. 

Our study also revealed the time-domain dynamical behavior of the system, showing that the dynamics of the system can be effectively reduced to a two-dimensional phase space defined by the amplitude and phase difference in the presence of gain saturation nonlinearity. This reduction implies that the system dynamics converge to a limit cycle in weak coupling regime. Moreover, the beat oscillations in linear APT-symmetric systems driven by the coherent energy exchange mechanism persist even in the nonlinear regime. Remarkably, even in the nonlinear regime above threshold, the oscillation frequencies coincide with the eigenfrequencies of the linear Hamiltonian, including the location and behavior of EPs, indicating a deep continuity across the linear and nonlinear domains.

Consequently, our study provides a new perspective on the relationship between EPs and nonlinear bifurcation points, highlighting intriguing links between linear non-Hermitian systems and nonlinear physics.

\begin{acknowledgments}
  We acknowledge fruitful discussions with Masaya Arahata.
  This work was supported by the Japan Society for the Promotion of Science (Grant number JP20H05641, JP21K14551, 24K01377, 24H02232, 24H00400 and JPMJSP2180).
\end{acknowledgments}

\section*{Data Availability}
Data underlying the results presented in this paper are not publicly available at this time but may be obtained from the authors upon reasonable request.

\appendix

\section{The parameters used in the rate equations}
\label{sec:rate_eq_params}

The parameters used in the rate equations Eqs. \eqref{eq:rate_eq_amp} and \eqref{eq:rate_eq_phase} are summarized in Table \ref{tab:rate_eq_params}. The values are chosen to be consistent with the experimental conditions of the APT-symmetric laser system \cite{Ji2023}.

\begin{table*}[htbp]
\centering
\begin{tabular}{l l l}
\hline
\hline
\textbf{Symbol} & \textbf{Parameter} & \textbf{Value} \\
\hline
$\omega_{1,2}$ & The frequencies of cavity 1,2 & $2\pi c / \lambda_{1,2}$ \\
$\lambda_{1,2}$ & The wavelengths of cavity 1,2 & $\lambda_0 \pm \Delta \lambda$ \\
$\lambda_0$ & The center wavelength & $1550$ [nm] \\
$\Delta \lambda$ & The wavelength detuning & $0.1$ [nm] \\
$\kappa$ & The cavity loss rate & $140.86$ [GHz] \\
$\beta$ & The spontaneous emission coefficient & $0.017$ \\
$\gamma_\parallel$ & The two-level radiative recombination rate & $2.2$ [GHz] \\
$\gamma_{\text{tot}}$ & The total carrier recombination rate & $5$ [GHz] \\
$V_a$ & The active volume & $0.016 \times 10^{-12}$ [cm$^3$] \\
$n_0$ & The carrier number at transparency & $10^{18}$ [cm$^{-3}] \times V_a = 16000$ \\
$P_0$ & The threshold of the single cavity laser & $\gamma_{\text{tot}} (n_0 + 2\kappa / (\beta \gamma_\parallel))$ \\
$P_\mathrm{tot}$ & The total pump rate & $3P_0$ \\
$P_{1,2}$ & The pump rate in cavity 1,2 & $0.5 P_\mathrm{tot}$ \\
\hline
\hline
\end{tabular}
\caption{
  Parameters used in the rate equations Eqs. \eqref{eq:rate_eq_amp} and \eqref{eq:rate_eq_phase}.} 
\label{tab:rate_eq_params}
\end{table*}

\section{Nonlinear equilibrium states in PT-symmetric coupled-cavity systems}
\label{sec:PT_w_StuartLandau}

\begin{figure}%
  \includegraphics[width=7cm]{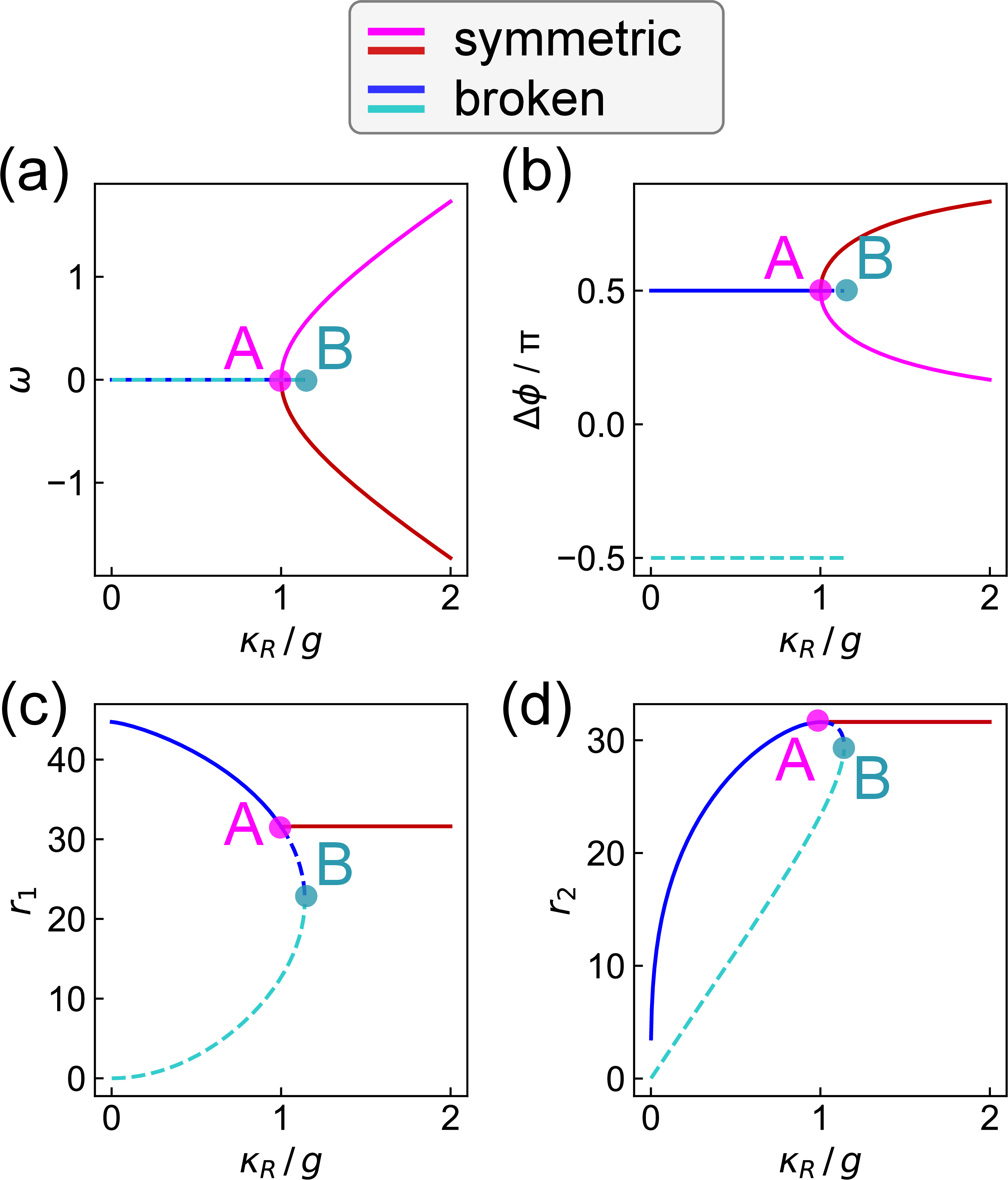}
  \caption{\label{figS1}
    Steady-state solutions of the PT-symmetric coupled-cavity system with gain saturation for $\gamma_{0}=1$, $\beta=10^{-3}$, and $g=1$, plotted as a function of the real coupling coefficient $\kappa_{R}$ in the range $0<\kappa_{R}<2$. 
    The panels show (a) oscillation frequency $\omega$, (b) phase difference $\Delta\phi$, and (c,d) amplitudes $r_{1}$ and $r_{2}$. 
    Solid lines denote stable equilibria, and dashed lines denote unstable equilibria. 
  }
\end{figure}

This section investigates the equilibrium states of a PT-symmetric coupled-cavity system with gain saturation. The system is described by a Stuart-Landau model, which is analogous to a class-A laser where carrier dynamics can be adiabatically eliminated. The Hamiltonian with Stuart-Landau nonlinearity is written as
\begin{align}
  H(\boldsymbol{a}) &= 
  \begin{bmatrix}
  \omega_0 + i(\gamma_1 - \beta{|a_1|}^2) & \kappa_R \\
  \kappa_R & \omega_0 + i(\gamma_2 - \beta{|a_2|}^2)
  \end{bmatrix}, \nonumber \\
  &\frac{\mathrm{d}}{\mathrm{d} t} \boldsymbol{a}(t) = -i\,H(\boldsymbol{a}(t))\,\boldsymbol{a}(t),
\end{align}
where $\gamma_1 = \gamma_0 + g$ and $\gamma_2 = \gamma_0 - g$. Here, $\omega_0$ is the resonance frequency, $\beta$ is the nonlinear coefficient, $\kappa_R$ is the real coupling coefficient, and $g$ denotes the gain difference between the two cavities. Using the polar representation $a_i = r_i e^{-i(\omega t + \phi_i)}$ $(i=1,2)$, and defining $\Delta \phi = \phi_2 - \phi_1$, we obtain

\begin{align}
\dot{r_1}  &= \bigl(\gamma_1-\beta r_1^{2}\bigr)\,r_1 - \kappa_R\,r_2 \sin\!\Delta\phi,\\
\dot{r_2} &= \bigl(\gamma_2-\beta r_2^{2}\bigr)\,r_2 + \kappa_R\,r_1 \sin\!\Delta\phi,\\
\dot{\phi_1} &= \bigl(\omega_0-\omega\bigr) + \kappa_R\,\frac{r_2}{r_1} \cos\!\Delta\phi,\\
\dot{\phi_2} &= \bigl(\omega_0-\omega\bigr) + \kappa_R\,\frac{r_1}{r_2} \cos\!\Delta\phi.
\end{align}

The steady-state conditions are $\dot{r}_{1,2} = 0$ and $\dot{\Delta\phi} = 0$. The latter yields
\begin{align}
\Delta\dot{\phi} &= \dot{\phi}_2 - \dot{\phi}_1
= \kappa_R\left(\frac{r_1}{r_2} - \frac{r_2}{r_1}\right)\cos\Delta\phi = 0,
\end{align}
implying either
\begin{align}
r_1 = r_2 =: r > 0, \quad \text{or} \quad \cos\Delta\phi = 0 \; \Rightarrow \; \Delta\phi = \pm\frac{\pi}{2}.
\end{align}
These correspond, respectively, to the PT-symmetric and PT-broken states in linear eigenvalue analysis. We will therefore refer to them as the symmetric and broken states.

For $r_1 = r_2 = r > 0$, substituting into $\dot{r}_{1,2} = 0$ yields

\begin{align}
\begin{cases}
(\gamma_1 - \beta r^{2})\,r - \kappa_R r\sin\Delta\phi = 0,\\
(\gamma_2 - \beta r^{2})\,r + \kappa_R r\sin\Delta\phi = 0,
\end{cases}
\end{align}

which give
\begin{align}
&\gamma_1 + \gamma_2 - 2\beta r^{2} = 0
  &&\Longrightarrow\quad r^{2} = \frac{\gamma_0}{\beta},\\[6pt]
&\gamma_1 - \gamma_2 - 2\kappa_R \sin\Delta\phi = 0
  &&\Longrightarrow\quad \sin\Delta\phi = \frac{g}{\kappa_R}.
\end{align}
The existence conditions are $\gamma_0 > 0$ and $|g| \le |\kappa_R|$.  
From the phase equations, the oscillation frequency is
\[
\omega = \omega_0 + \kappa_R\cos\Delta\phi = \omega_0 \pm \sqrt{\kappa_R^2 - g^2}.
\]

For $\Delta\phi = \pm\frac{\pi}{2}$, let $\sin\Delta\phi = s = \pm 1$. Then
\begin{align}
(\gamma_1 - \beta r_1^2) r_1 &= s\,\kappa_R r_2,\\
(\gamma_2 - \beta r_2^2) r_2 &= -s\,\kappa_R r_1.
\end{align}
Multiplying these equations gives
\begin{align}
(\gamma_1 - \beta r_1^2)(\gamma_2 - \beta r_2^2) &= -\kappa_R^2, \\
\gamma_1 r_1^2 + \gamma_2 r_2^2 &= \beta(r_1^4 + r_2^4).
\end{align}
Introducing $u := \gamma_1 - \beta r_1^2$, we have
\[
r_1^2 = \frac{\gamma_1 - u}{\beta},\quad
r_2^2 = \frac{\gamma_2 + \kappa_R^2/u}{\beta},
\]
where $u \neq 0$ satisfies the quartic equation
\[
u^4 - (\gamma_0 + g)u^3 + (\gamma_0 - g)\kappa_R^2 u + \kappa_R^4 = 0.
\]

The formula for the oscillation frequency is derived from $\cos\Delta\phi = 0$.
\[
\omega = \omega_0.
\]

Figure~\ref{figS1} plots $\omega$, $\Delta\phi$, $r_1$, and $r_2$ versus $\kappa_R$ for $\gamma_0 = 1$, $\beta = 10^{-3}$, and $g = 1$ in the range $0 < \kappa_R < 2$. Solid curves denote stable equilibria and dashed curves denote unstable ones.  
Point~A corresponds to a pitchfork bifurcation where two stable symmetric branches and one unstable branch merge. Point~B corresponds to a saddle-node bifurcation where a stable broken branch and an unstable symmetric branch merge into a single stable symmetric branch. As in APT-symmetric systems, at Point~A we have $r_1 = r_2$ and $\Delta\phi = \pi/2$, and the Hamiltonian becomes defective. For $\kappa_R < g$, only one of the broken-state solutions is stable, leading to single-mode oscillation. For $\kappa_R \ge g$, solutions that inherit the original PT symmetry emerge and are both stable, resulting in bistability. Thus, point~A marks the boundary between monostability and the bistable regime. The stable EP-derived bifurcation point is a feature of this simplified class-A laser model. As discussed in the main text and shown in Ref. \cite{Ji2023}, this point becomes unstable in a class-B laser model where carrier dynamics are included, typically leading to sustained relaxation oscillations.

\section{Convergence of the amplitude difference of the two cavities under $\kappa_I < \Delta \omega$}
\label{sec:appendix_convergence}

The ``broken'' equilibrium states in Eq. (11) in the main text are characterized by \( r_1 \neq r_2 \); however, this state is not observed in numerical simulations of time evolution. Instead, the system converges to a steady state with the synchronized amplitudes \( r_1 = r_2 \). Furthermore, when \( \kappa_I < \Delta\omega \), the condition \( \mathrm{d}\Delta\phi / \mathrm{d}t \neq 0 \) holds, leading to the appearance of self-pulsations in the amplitude of the system. This behavior is observed in the upper panel of Fig. 3 in the main text.
In the following discussion, we analyze the dynamics that are observed in the steady state under the condition \( \kappa_I < \Delta\omega \), and show that the amplitude of the two cavities converges to the same value, \( r_1 = r_2 \), and that the beat frequency is consistent with the analytical prediction derived under \( r_1 = r_2 \).
%
\begin{figure}%
  \includegraphics[width=8cm]{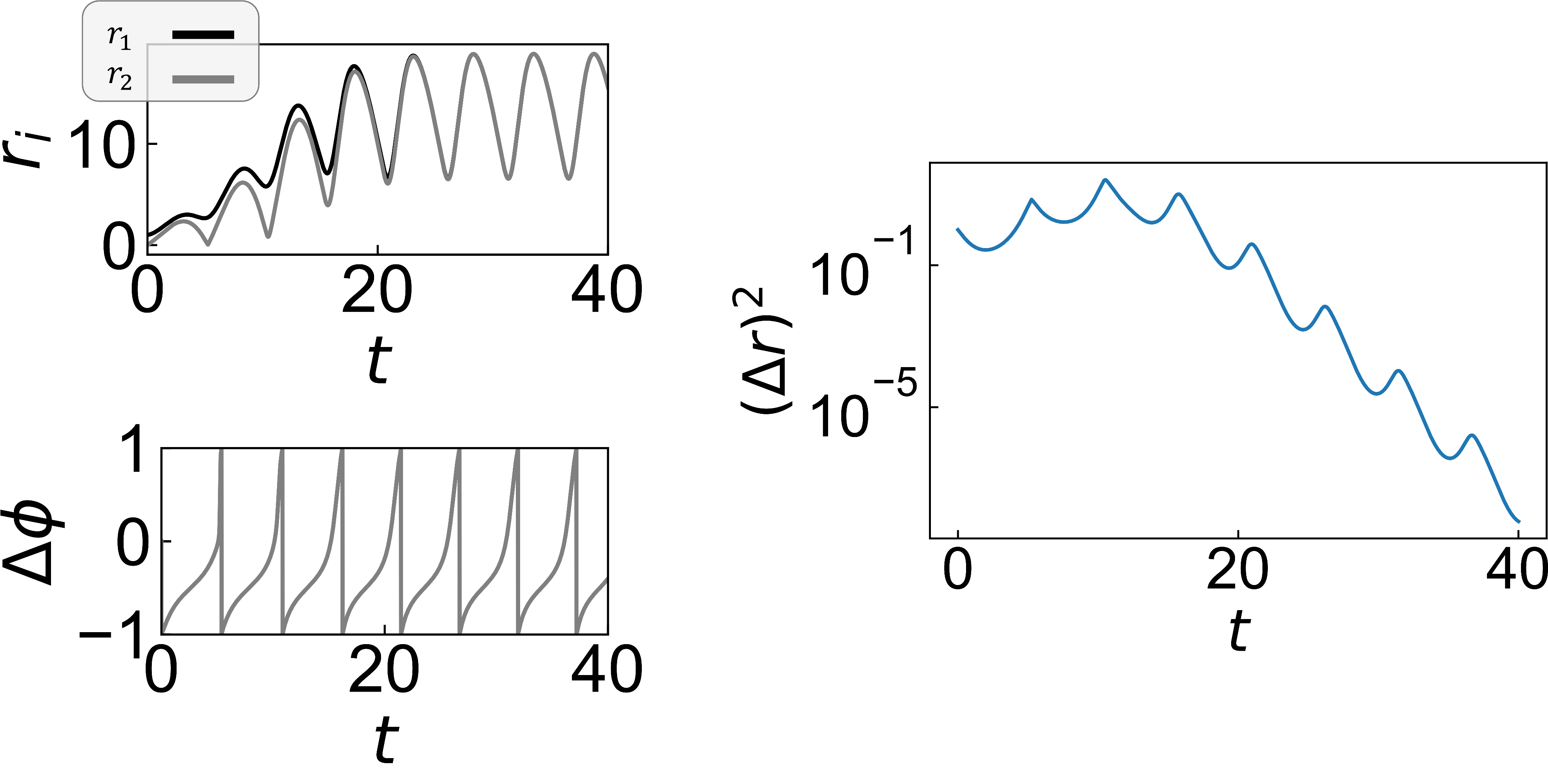}
  \caption{\label{figS2}
		Typical time evolution of the amplitudes $r_1, r_2$, phase difference $\Delta \phi$ and the difference of the amplitude $\Delta r$ of the system.
		The coupling strength are set to $\kappa_I = 0.8 \Delta\omega$, and the other parameters are the same as those in the main text.
  }
\end{figure}

Figure \ref{figS2} shows the typical time evolution of the system. The coupling strength is set to \( \kappa_I = 0.8 \Delta\omega \) (\( < \Delta \omega \)). 
As shown in Fig. \ref{figS2}, the system transitions to a steady state through the following two sequential processes:
\begin{enumerate}
	\item Saturation of \( r_1 \) and \( r_2 \) due to the effects of third-order nonlinearity.
	\item Convergence of \( \Delta r := r_2 - r_1 \) to zero.
\end{enumerate}
The first process can be readily understood as a natural consequence of the gain saturation. Therefore, the primary objective of this section is to formulate the second process and to verify whether the dynamics of our system at \( \kappa_I < \Delta\omega \) can be explained by assuming \( \Delta r \to 0 \).

We start with the equations of motion in the main text, Eqs. \eqref{eq:nonlinear_r1} to \eqref{eq:nonlinear_phi2}. By subtracting Eq. (\ref{eq:nonlinear_r1}) from Eq. (\ref{eq:nonlinear_r2}) and we get the equation of motion for \( \Delta r \):
\begin{align}
	\label{eq:nonlinear_dr_full}
	\frac{\mathrm{d} }{\mathrm{d} t} \Delta r 
	&= \left(\gamma - \kappa_I \cos \Delta \phi - \left( r_1^2 + r_1 r_2 + r_2^2 \right) \right) \Delta r 
\end{align}
From \eqref{eq:nonlinear_dr_full}, the sufficient conditions for $\Delta r \to 0$ are as follows.
\begin{align}
	\label{eq:convergence_condition_dr}
	r_1^2 + r_1 r_2 + r_2^2 > \frac{\gamma - \kappa_I \cos \Delta \phi }{ \beta }
\end{align}
In fact, even when \eqref{eq:convergence_condition_dr} is not satisfied, $\Delta r \to 0$ as $t \to \infty$ holds for general $\kappa_I$ and $\gamma$. 
By evaluating only the first order terms for $\Delta r$ of \eqref{eq:nonlinear_dr_full}, we obtain
\begin{align}
	\label{eq:nonlinear_dr}
	\frac{\mathrm{d} }{\mathrm{d} t} \Delta r &= (\gamma - \kappa_I \cos \Delta \phi - 3\beta \gamma r_1^2) \Delta r + \mathcal{O} \left( {(\Delta r)}^2 \right)
\end{align}
In the following, some approximations are made to simplify the analysis.
First, $r_1 \sim \sqrt{\gamma/\beta}$ roughly holds when $r_1, r_2$ are saturated.
This approximation is valid when the coupling strength $\kappa_I$ is weak, where the two cavities are almost independent of each other.
In this situation, \eqref{eq:nonlinear_dr} can be approximated as
\begin{align}
	\label{eq:nonlinear_dr_approx}
	\frac{\mathrm{d} }{\mathrm{d} t} \Delta r &\sim (-2\gamma - \kappa_I \cos \Delta \phi ) \Delta r 
\end{align}
In addition, the net gain $\gamma$ is larger than the coupling strength $\kappa_I$, 
the oscillation speed $\Delta \phi$ is fast, 
\begin{align}
	\label{eq:nonlinear_dr_approx2}
	\frac{\mathrm{d} }{\mathrm{d} t} \Delta r \sim -2\gamma \Delta r 
\end{align}
holds on average. 
The approximation in Eq. (\ref{eq:nonlinear_dr_approx}) is not valid when either of the following conditions is satisfied:
\begin{enumerate}
	\item The coupling strength $\kappa_I$ is not weak ($\kappa_I \sim \Delta\omega$).
	\item The net gain $\gamma$ is not larger than the coupling strength $\kappa_I$ ($\gamma \ll \kappa_I$).
\end{enumerate}
Therefore, \eqref{eq:nonlinear_dr_approx2} cannot directly predict the converges of $\Delta r$ to zero where the coupling is relatively large and the net gain is small (\( \kappa_I \sim \Delta\omega, \gamma \ll 1 \)).
In both cases, however, we can see that $\Delta r$ converges to zero as $t\to\infty$.
%
\begin{figure}%
  \includegraphics[width=8cm]{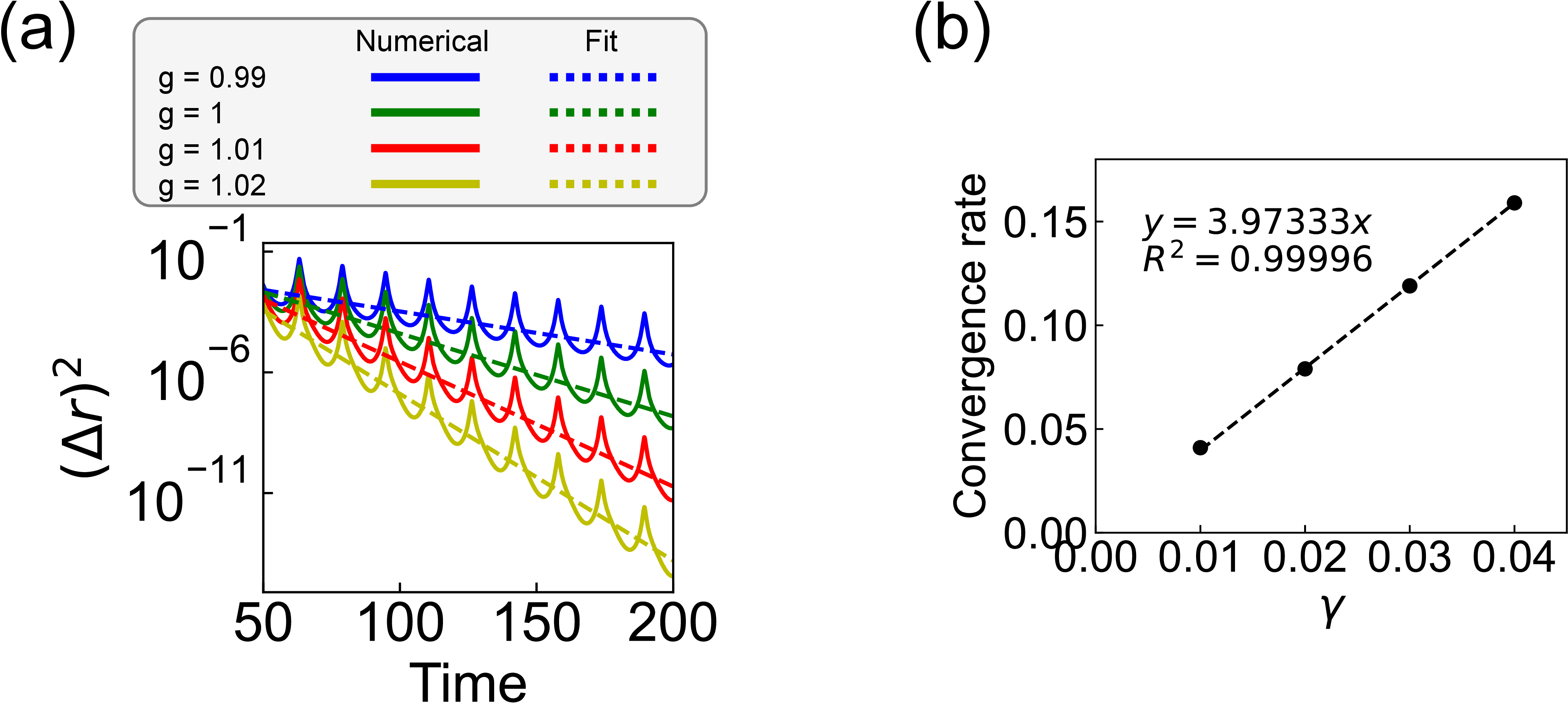}
  \caption{\label{figS3}
		(a) Time evolution of ${(\Delta r)}$ for $\kappa_I = 0.49$ and $g = 0.99, 1, 1.01, 1.02$.
		(b) The slope of the linear fit to ${(\Delta r)}^2$ for $t > 50$ as a function of net gain $\gamma$.
  }
\end{figure}

Figure \ref{figS3} (a) shows the time evolution of \( \Delta r \) for \( \kappa_I = 0.49 \), with the gain set to four different values: \( g = 0.99, 1, 1.01, 1.02 \) (\( \gamma = 0.01, 0.02, 0.03, 0.04 \)). The linear fit is applied to the data for \( t > 50 \), after significant gain saturation occurs.
In this parameter setting, \( \gamma \ll \kappa_I \), and the beat period is larger compared to the intrinsic oscillation timescale (\( \omega \sim 1 \)). Thus, this situation satisfies neither condition (i) nor (ii). Nevertheless, the slope of the approximate linear fit, which corresponds to the convergence rate of \( (\Delta r)^2 \), is proportional to \( \gamma \), as shown in Fig. \ref{figS3}(b). This strongly suggests that \eqref{eq:nonlinear_dr_approx2} holds even in conditions where neither (i) nor (ii) is valid. As indicated by the results in Fig. \ref{figS3}, the time-averaged contribution of the $\cos \Delta \phi$ term becomes zero. However, proving this requires relying on numerical methods.

\bibliography{reference}

\begin{thebibliography}{42}%
\makeatletter
\providecommand \@ifxundefined [1]{%
 \@ifx{#1\undefined}
}%
\providecommand \@ifnum [1]{%
 \ifnum #1\expandafter \@firstoftwo
 \else \expandafter \@secondoftwo
 \fi
}%
\providecommand \@ifx [1]{%
 \ifx #1\expandafter \@firstoftwo
 \else \expandafter \@secondoftwo
 \fi
}%
\providecommand \natexlab [1]{#1}%
\providecommand \enquote  [1]{``#1''}%
\providecommand \bibnamefont  [1]{#1}%
\providecommand \bibfnamefont [1]{#1}%
\providecommand \citenamefont [1]{#1}%
\providecommand \href@noop [0]{\@secondoftwo}%
\providecommand \href [0]{\begingroup \@sanitize@url \@href}%
\providecommand \@href[1]{\@@startlink{#1}\@@href}%
\providecommand \@@href[1]{\endgroup#1\@@endlink}%
\providecommand \@sanitize@url [0]{\catcode `\\12\catcode `\$12\catcode `\&12\catcode `\#12\catcode `\^12\catcode `\_12\catcode `\%12\relax}%
\providecommand \@@startlink[1]{}%
\providecommand \@@endlink[0]{}%
\providecommand \url  [0]{\begingroup\@sanitize@url \@url }%
\providecommand \@url [1]{\endgroup\@href {#1}{\urlprefix }}%
\providecommand \urlprefix  [0]{URL }%
\providecommand \Eprint [0]{\href }%
\providecommand \doibase [0]{https://doi.org/}%
\providecommand \selectlanguage [0]{\@gobble}%
\providecommand \bibinfo  [0]{\@secondoftwo}%
\providecommand \bibfield  [0]{\@secondoftwo}%
\providecommand \translation [1]{[#1]}%
\providecommand \BibitemOpen [0]{}%
\providecommand \bibitemStop [0]{}%
\providecommand \bibitemNoStop [0]{.\EOS\space}%
\providecommand \EOS [0]{\spacefactor3000\relax}%
\providecommand \BibitemShut  [1]{\csname bibitem#1\endcsname}%
\let\auto@bib@innerbib\@empty
\bibitem [{\citenamefont {Bender}\ and\ \citenamefont {Boettcher}(1998)}]{PhysRevLett.80.5243}%
  \BibitemOpen
  \bibfield  {author} {\bibinfo {author} {\bibfnamefont {C.~M.}\ \bibnamefont {Bender}}\ and\ \bibinfo {author} {\bibfnamefont {S.}~\bibnamefont {Boettcher}},\ }\bibfield  {title} {\bibinfo {title} {Real spectra in non-hermitian hamiltonians having $\mathcal{P}\mathcal{T}$ symmetry},\ }\href {https://doi.org/10.1103/PhysRevLett.80.5243} {\bibfield  {journal} {\bibinfo  {journal} {Phys. Rev. Lett.}\ }\textbf {\bibinfo {volume} {80}},\ \bibinfo {pages} {5243} (\bibinfo {year} {1998})}\BibitemShut {NoStop}%
\bibitem [{\citenamefont {Bender}(2007)}]{Bender_2007}%
  \BibitemOpen
  \bibfield  {author} {\bibinfo {author} {\bibfnamefont {C.~M.}\ \bibnamefont {Bender}},\ }\bibfield  {title} {\bibinfo {title} {Making sense of non-hermitian hamiltonians},\ }\href {https://doi.org/10.1088/0034-4885/70/6/R03} {\bibfield  {journal} {\bibinfo  {journal} {Reports on Progress in Physics}\ }\textbf {\bibinfo {volume} {70}},\ \bibinfo {pages} {947} (\bibinfo {year} {2007})}\BibitemShut {NoStop}%
\bibitem [{\citenamefont {Regensburger}\ \emph {et~al.}(2012)\citenamefont {Regensburger}, \citenamefont {Bersch}, \citenamefont {Miri}, \citenamefont {Onishchukov}, \citenamefont {Christodoulides},\ and\ \citenamefont {Peschel}}]{Regensburger2012}%
  \BibitemOpen
  \bibfield  {author} {\bibinfo {author} {\bibfnamefont {A.}~\bibnamefont {Regensburger}}, \bibinfo {author} {\bibfnamefont {C.}~\bibnamefont {Bersch}}, \bibinfo {author} {\bibfnamefont {M.-A.}\ \bibnamefont {Miri}}, \bibinfo {author} {\bibfnamefont {G.}~\bibnamefont {Onishchukov}}, \bibinfo {author} {\bibfnamefont {D.~N.}\ \bibnamefont {Christodoulides}},\ and\ \bibinfo {author} {\bibfnamefont {U.}~\bibnamefont {Peschel}},\ }\bibfield  {title} {\bibinfo {title} {Parity-time synthetic photonic lattices},\ }\href {https://doi.org/10.1038/nature11298} {\bibfield  {journal} {\bibinfo  {journal} {Nature}\ }\textbf {\bibinfo {volume} {488}},\ \bibinfo {pages} {167} (\bibinfo {year} {2012})}\BibitemShut {NoStop}%
\bibitem [{\citenamefont {Feng}\ \emph {et~al.}(2017)\citenamefont {Feng}, \citenamefont {El-Ganainy},\ and\ \citenamefont {Ge}}]{Feng2017}%
  \BibitemOpen
  \bibfield  {author} {\bibinfo {author} {\bibfnamefont {L.}~\bibnamefont {Feng}}, \bibinfo {author} {\bibfnamefont {R.}~\bibnamefont {El-Ganainy}},\ and\ \bibinfo {author} {\bibfnamefont {L.}~\bibnamefont {Ge}},\ }\bibfield  {title} {\bibinfo {title} {Non-hermitian photonics based on parity--time symmetry},\ }\href {https://doi.org/10.1038/s41566-017-0031-1} {\bibfield  {journal} {\bibinfo  {journal} {Nature Photonics}\ }\textbf {\bibinfo {volume} {11}},\ \bibinfo {pages} {752} (\bibinfo {year} {2017})}\BibitemShut {NoStop}%
\bibitem [{\citenamefont {{\"O}zdemir}\ \emph {et~al.}(2019)\citenamefont {{\"O}zdemir}, \citenamefont {Rotter}, \citenamefont {Nori},\ and\ \citenamefont {Yang}}]{Ozdemir2019}%
  \BibitemOpen
  \bibfield  {author} {\bibinfo {author} {\bibfnamefont {{\c{S}}.~K.}\ \bibnamefont {{\"O}zdemir}}, \bibinfo {author} {\bibfnamefont {S.}~\bibnamefont {Rotter}}, \bibinfo {author} {\bibfnamefont {F.}~\bibnamefont {Nori}},\ and\ \bibinfo {author} {\bibfnamefont {L.}~\bibnamefont {Yang}},\ }\bibfield  {title} {\bibinfo {title} {Parity--time symmetry and exceptional points in photonics},\ }\href {https://doi.org/10.1038/s41563-019-0304-9} {\bibfield  {journal} {\bibinfo  {journal} {Nature Materials}\ }\textbf {\bibinfo {volume} {18}},\ \bibinfo {pages} {783} (\bibinfo {year} {2019})}\BibitemShut {NoStop}%
\bibitem [{\citenamefont {Miri}\ and\ \citenamefont {Al^^c3^^b9}(2019)}]{doi:10.1126/science.aar7709}%
  \BibitemOpen
  \bibfield  {author} {\bibinfo {author} {\bibfnamefont {M.-A.}\ \bibnamefont {Miri}}\ and\ \bibinfo {author} {\bibfnamefont {A.}~\bibnamefont {Al^^c3^^b9}},\ }\bibfield  {title} {\bibinfo {title} {Exceptional points in optics and photonics},\ }\href {https://doi.org/10.1126/science.aar7709} {\bibfield  {journal} {\bibinfo  {journal} {Science}\ }\textbf {\bibinfo {volume} {363}},\ \bibinfo {pages} {eaar7709} (\bibinfo {year} {2019})}\BibitemShut {NoStop}%
\bibitem [{\citenamefont {Choi}\ \emph {et~al.}(2018)\citenamefont {Choi}, \citenamefont {Hahn}, \citenamefont {Yoon},\ and\ \citenamefont {Song}}]{Choi2018}%
  \BibitemOpen
  \bibfield  {author} {\bibinfo {author} {\bibfnamefont {Y.}~\bibnamefont {Choi}}, \bibinfo {author} {\bibfnamefont {C.}~\bibnamefont {Hahn}}, \bibinfo {author} {\bibfnamefont {J.~W.}\ \bibnamefont {Yoon}},\ and\ \bibinfo {author} {\bibfnamefont {S.~H.}\ \bibnamefont {Song}},\ }\bibfield  {title} {\bibinfo {title} {Observation of an anti-pt-symmetric exceptional point and energy-difference conserving dynamics in electrical circuit resonators},\ }\href {https://doi.org/10.1038/s41467-018-04690-y} {\bibfield  {journal} {\bibinfo  {journal} {Nature Communications}\ }\textbf {\bibinfo {volume} {9}},\ \bibinfo {pages} {2182} (\bibinfo {year} {2018})}\BibitemShut {NoStop}%
\bibitem [{\citenamefont {Zhang}\ \emph {et~al.}(2020)\citenamefont {Zhang}, \citenamefont {Huang}, \citenamefont {Zhang}, \citenamefont {Li}, \citenamefont {Qiu}, \citenamefont {Nori},\ and\ \citenamefont {Jing}}]{Zhang2020}%
  \BibitemOpen
  \bibfield  {author} {\bibinfo {author} {\bibfnamefont {H.}~\bibnamefont {Zhang}}, \bibinfo {author} {\bibfnamefont {R.}~\bibnamefont {Huang}}, \bibinfo {author} {\bibfnamefont {S.-D.}\ \bibnamefont {Zhang}}, \bibinfo {author} {\bibfnamefont {Y.}~\bibnamefont {Li}}, \bibinfo {author} {\bibfnamefont {C.-W.}\ \bibnamefont {Qiu}}, \bibinfo {author} {\bibfnamefont {F.}~\bibnamefont {Nori}},\ and\ \bibinfo {author} {\bibfnamefont {H.}~\bibnamefont {Jing}},\ }\bibfield  {title} {\bibinfo {title} {Breaking anti-pt symmetry by spinning a resonator},\ }\href {https://doi.org/10.1021/acs.nanolett.0c03119} {\bibfield  {journal} {\bibinfo  {journal} {Nano Letters}\ }\textbf {\bibinfo {volume} {20}},\ \bibinfo {pages} {7594} (\bibinfo {year} {2020})}\BibitemShut {NoStop}%
\bibitem [{\citenamefont {Li}\ \emph {et~al.}(2024)\citenamefont {Li}, \citenamefont {Chen}, \citenamefont {Wu}, \citenamefont {Wang}, \citenamefont {Wang}, \citenamefont {Zhong}, \citenamefont {Huang}, \citenamefont {Liu}, \citenamefont {Chen}, \citenamefont {Luo},\ and\ \citenamefont {Chen}}]{Li2024}%
  \BibitemOpen
  \bibfield  {author} {\bibinfo {author} {\bibfnamefont {H.}~\bibnamefont {Li}}, \bibinfo {author} {\bibfnamefont {L.}~\bibnamefont {Chen}}, \bibinfo {author} {\bibfnamefont {W.}~\bibnamefont {Wu}}, \bibinfo {author} {\bibfnamefont {H.}~\bibnamefont {Wang}}, \bibinfo {author} {\bibfnamefont {T.}~\bibnamefont {Wang}}, \bibinfo {author} {\bibfnamefont {Y.}~\bibnamefont {Zhong}}, \bibinfo {author} {\bibfnamefont {F.}~\bibnamefont {Huang}}, \bibinfo {author} {\bibfnamefont {G.-S.}\ \bibnamefont {Liu}}, \bibinfo {author} {\bibfnamefont {Y.}~\bibnamefont {Chen}}, \bibinfo {author} {\bibfnamefont {Y.}~\bibnamefont {Luo}},\ and\ \bibinfo {author} {\bibfnamefont {Z.}~\bibnamefont {Chen}},\ }\bibfield  {title} {\bibinfo {title} {Enhanced sensitivity with nonlinearity-induced exceptional points degeneracy lifting},\ }\href {https://doi.org/10.1038/s42005-024-01609-6} {\bibfield  {journal} {\bibinfo  {journal} {Communications Physics}\ }\textbf {\bibinfo {volume} {7}},\ \bibinfo {pages} {117} (\bibinfo {year} {2024})}\BibitemShut {NoStop}%
\bibitem [{\citenamefont {Heiss}\ and\ \citenamefont {Harney}(2001)}]{Heiss2001}%
  \BibitemOpen
  \bibfield  {author} {\bibinfo {author} {\bibfnamefont {W.~D.}\ \bibnamefont {Heiss}}\ and\ \bibinfo {author} {\bibfnamefont {H.~L.}\ \bibnamefont {Harney}},\ }\bibfield  {title} {\bibinfo {title} {The chirality of exceptional points},\ }\href {https://doi.org/10.1007/s100530170017} {\bibfield  {journal} {\bibinfo  {journal} {The European Physical Journal D - Atomic, Molecular, Optical and Plasma Physics}\ }\textbf {\bibinfo {volume} {17}},\ \bibinfo {pages} {149} (\bibinfo {year} {2001})}\BibitemShut {NoStop}%
\bibitem [{\citenamefont {Takata}\ and\ \citenamefont {Notomi}(2017)}]{PhysRevApplied.7.054023}%
  \BibitemOpen
  \bibfield  {author} {\bibinfo {author} {\bibfnamefont {K.}~\bibnamefont {Takata}}\ and\ \bibinfo {author} {\bibfnamefont {M.}~\bibnamefont {Notomi}},\ }\bibfield  {title} {\bibinfo {title} {$\mathcal{P}\mathcal{T}$-symmetric coupled-resonator waveguide based on buried heterostructure nanocavities},\ }\href {https://doi.org/10.1103/PhysRevApplied.7.054023} {\bibfield  {journal} {\bibinfo  {journal} {Phys. Rev. Appl.}\ }\textbf {\bibinfo {volume} {7}},\ \bibinfo {pages} {054023} (\bibinfo {year} {2017})}\BibitemShut {NoStop}%
\bibitem [{\citenamefont {Peng}\ \emph {et~al.}(2014)\citenamefont {Peng}, \citenamefont {{\"O}zdemir}, \citenamefont {Lei}, \citenamefont {Monifi}, \citenamefont {Gianfreda}, \citenamefont {Long}, \citenamefont {Fan}, \citenamefont {Nori}, \citenamefont {Bender},\ and\ \citenamefont {Yang}}]{Peng2014}%
  \BibitemOpen
  \bibfield  {author} {\bibinfo {author} {\bibfnamefont {B.}~\bibnamefont {Peng}}, \bibinfo {author} {\bibfnamefont {{\c{S}}.~K.}\ \bibnamefont {{\"O}zdemir}}, \bibinfo {author} {\bibfnamefont {F.}~\bibnamefont {Lei}}, \bibinfo {author} {\bibfnamefont {F.}~\bibnamefont {Monifi}}, \bibinfo {author} {\bibfnamefont {M.}~\bibnamefont {Gianfreda}}, \bibinfo {author} {\bibfnamefont {G.~L.}\ \bibnamefont {Long}}, \bibinfo {author} {\bibfnamefont {S.}~\bibnamefont {Fan}}, \bibinfo {author} {\bibfnamefont {F.}~\bibnamefont {Nori}}, \bibinfo {author} {\bibfnamefont {C.~M.}\ \bibnamefont {Bender}},\ and\ \bibinfo {author} {\bibfnamefont {L.}~\bibnamefont {Yang}},\ }\bibfield  {title} {\bibinfo {title} {Parity--time-symmetric whispering-gallery microcavities},\ }\href {https://doi.org/10.1038/nphys2927} {\bibfield  {journal} {\bibinfo  {journal} {Nature Physics}\ }\textbf {\bibinfo {volume} {10}},\ \bibinfo {pages} {394} (\bibinfo {year} {2014})}\BibitemShut {NoStop}%
\bibitem [{\citenamefont {Peng}\ \emph {et~al.}(2016)\citenamefont {Peng}, \citenamefont {^^c5^^9eahin Kaya~^^c3^^96zdemir}, \citenamefont {Liertzer}, \citenamefont {Chen}, \citenamefont {Kramer}, \citenamefont {Y^^c4^^b1lmaz}, \citenamefont {Wiersig}, \citenamefont {Rotter},\ and\ \citenamefont {Yang}}]{doi:10.1073/pnas.1603318113}%
  \BibitemOpen
  \bibfield  {author} {\bibinfo {author} {\bibfnamefont {B.}~\bibnamefont {Peng}}, \bibinfo {author} {\bibnamefont {^^c5^^9eahin Kaya~^^c3^^96zdemir}}, \bibinfo {author} {\bibfnamefont {M.}~\bibnamefont {Liertzer}}, \bibinfo {author} {\bibfnamefont {W.}~\bibnamefont {Chen}}, \bibinfo {author} {\bibfnamefont {J.}~\bibnamefont {Kramer}}, \bibinfo {author} {\bibfnamefont {H.}~\bibnamefont {Y^^c4^^b1lmaz}}, \bibinfo {author} {\bibfnamefont {J.}~\bibnamefont {Wiersig}}, \bibinfo {author} {\bibfnamefont {S.}~\bibnamefont {Rotter}},\ and\ \bibinfo {author} {\bibfnamefont {L.}~\bibnamefont {Yang}},\ }\bibfield  {title} {\bibinfo {title} {Chiral modes and directional lasing at exceptional points},\ }\href {https://doi.org/10.1073/pnas.1603318113} {\bibfield  {journal} {\bibinfo  {journal} {Proceedings of the National Academy of Sciences}\ }\textbf {\bibinfo {volume} {113}},\ \bibinfo {pages} {6845} (\bibinfo {year} {2016})}\BibitemShut {NoStop}%
\bibitem [{\citenamefont {Park}\ \emph {et~al.}(2020)\citenamefont {Park}, \citenamefont {Ndao}, \citenamefont {Cai}, \citenamefont {Hsu}, \citenamefont {Kodigala}, \citenamefont {Lepetit}, \citenamefont {Lo},\ and\ \citenamefont {Kant{\'e}}}]{Park2020}%
  \BibitemOpen
  \bibfield  {author} {\bibinfo {author} {\bibfnamefont {J.-H.}\ \bibnamefont {Park}}, \bibinfo {author} {\bibfnamefont {A.}~\bibnamefont {Ndao}}, \bibinfo {author} {\bibfnamefont {W.}~\bibnamefont {Cai}}, \bibinfo {author} {\bibfnamefont {L.}~\bibnamefont {Hsu}}, \bibinfo {author} {\bibfnamefont {A.}~\bibnamefont {Kodigala}}, \bibinfo {author} {\bibfnamefont {T.}~\bibnamefont {Lepetit}}, \bibinfo {author} {\bibfnamefont {Y.-H.}\ \bibnamefont {Lo}},\ and\ \bibinfo {author} {\bibfnamefont {B.}~\bibnamefont {Kant{\'e}}},\ }\bibfield  {title} {\bibinfo {title} {Symmetry-breaking-induced plasmonic exceptional points and nanoscale sensing},\ }\href {https://doi.org/10.1038/s41567-020-0796-x} {\bibfield  {journal} {\bibinfo  {journal} {Nature Physics}\ }\textbf {\bibinfo {volume} {16}},\ \bibinfo {pages} {462} (\bibinfo {year} {2020})}\BibitemShut {NoStop}%
\bibitem [{\citenamefont {Duggan}\ \emph {et~al.}(2022)\citenamefont {Duggan}, \citenamefont {Mann},\ and\ \citenamefont {Al{\`u}}}]{Duggan2022}%
  \BibitemOpen
  \bibfield  {author} {\bibinfo {author} {\bibfnamefont {R.}~\bibnamefont {Duggan}}, \bibinfo {author} {\bibfnamefont {S.~A.}\ \bibnamefont {Mann}},\ and\ \bibinfo {author} {\bibfnamefont {A.}~\bibnamefont {Al{\`u}}},\ }\bibfield  {title} {\bibinfo {title} {Limitations of sensing at an exceptional point},\ }\href {https://doi.org/10.1021/acsphotonics.1c01535} {\bibfield  {journal} {\bibinfo  {journal} {ACS Photonics}\ }\textbf {\bibinfo {volume} {9}},\ \bibinfo {pages} {1554} (\bibinfo {year} {2022})}\BibitemShut {NoStop}%
\bibitem [{\citenamefont {Kononchuk}\ \emph {et~al.}(2022)\citenamefont {Kononchuk}, \citenamefont {Cai}, \citenamefont {Ellis}, \citenamefont {Thevamaran},\ and\ \citenamefont {Kottos}}]{Kononchuk2022}%
  \BibitemOpen
  \bibfield  {author} {\bibinfo {author} {\bibfnamefont {R.}~\bibnamefont {Kononchuk}}, \bibinfo {author} {\bibfnamefont {J.}~\bibnamefont {Cai}}, \bibinfo {author} {\bibfnamefont {F.}~\bibnamefont {Ellis}}, \bibinfo {author} {\bibfnamefont {R.}~\bibnamefont {Thevamaran}},\ and\ \bibinfo {author} {\bibfnamefont {T.}~\bibnamefont {Kottos}},\ }\bibfield  {title} {\bibinfo {title} {Exceptional-point-based accelerometers with enhanced signal-to-noise ratio},\ }\href {https://doi.org/10.1038/s41586-022-04904-w} {\bibfield  {journal} {\bibinfo  {journal} {Nature}\ }\textbf {\bibinfo {volume} {607}},\ \bibinfo {pages} {697} (\bibinfo {year} {2022})}\BibitemShut {NoStop}%
\bibitem [{\citenamefont {Takata}\ \emph {et~al.}(2021)\citenamefont {Takata}, \citenamefont {Nozaki}, \citenamefont {Kuramochi}, \citenamefont {Matsuo}, \citenamefont {Takeda}, \citenamefont {Fujii}, \citenamefont {Kita}, \citenamefont {Shinya},\ and\ \citenamefont {Notomi}}]{Takata:21}%
  \BibitemOpen
  \bibfield  {author} {\bibinfo {author} {\bibfnamefont {K.}~\bibnamefont {Takata}}, \bibinfo {author} {\bibfnamefont {K.}~\bibnamefont {Nozaki}}, \bibinfo {author} {\bibfnamefont {E.}~\bibnamefont {Kuramochi}}, \bibinfo {author} {\bibfnamefont {S.}~\bibnamefont {Matsuo}}, \bibinfo {author} {\bibfnamefont {K.}~\bibnamefont {Takeda}}, \bibinfo {author} {\bibfnamefont {T.}~\bibnamefont {Fujii}}, \bibinfo {author} {\bibfnamefont {S.}~\bibnamefont {Kita}}, \bibinfo {author} {\bibfnamefont {A.}~\bibnamefont {Shinya}},\ and\ \bibinfo {author} {\bibfnamefont {M.}~\bibnamefont {Notomi}},\ }\bibfield  {title} {\bibinfo {title} {Observing exceptional point degeneracy of radiation with electrically pumped photonic crystal coupled-nanocavity lasers},\ }\href {https://doi.org/10.1364/OPTICA.412596} {\bibfield  {journal} {\bibinfo  {journal} {Optica}\ }\textbf {\bibinfo {volume} {8}},\ \bibinfo {pages} {184} (\bibinfo {year} {2021})}\BibitemShut {NoStop}%
\bibitem [{\citenamefont {Strogatz}(2018)}]{strogatz2018nonlinear}%
  \BibitemOpen
  \bibfield  {author} {\bibinfo {author} {\bibfnamefont {S.~H.}\ \bibnamefont {Strogatz}},\ }\href@noop {} {\emph {\bibinfo {title} {Nonlinear dynamics and chaos: with applications to physics, biology, chemistry, and engineering}}}\ (\bibinfo  {publisher} {CRC press},\ \bibinfo {year} {2018})\BibitemShut {NoStop}%
\bibitem [{\citenamefont {Bai}\ \emph {et~al.}(2022)\citenamefont {Bai}, \citenamefont {Fang}, \citenamefont {Liu}, \citenamefont {Li}, \citenamefont {Wan},\ and\ \citenamefont {Xiao}}]{10.1093/nsr/nwac259}%
  \BibitemOpen
  \bibfield  {author} {\bibinfo {author} {\bibfnamefont {K.}~\bibnamefont {Bai}}, \bibinfo {author} {\bibfnamefont {L.}~\bibnamefont {Fang}}, \bibinfo {author} {\bibfnamefont {T.-R.}\ \bibnamefont {Liu}}, \bibinfo {author} {\bibfnamefont {J.-Z.}\ \bibnamefont {Li}}, \bibinfo {author} {\bibfnamefont {D.}~\bibnamefont {Wan}},\ and\ \bibinfo {author} {\bibfnamefont {M.}~\bibnamefont {Xiao}},\ }\bibfield  {title} {\bibinfo {title} {{Nonlinearity-enabled higher-order exceptional singularities with ultra-enhanced signal-to-noise ratio}},\ }\href {https://doi.org/10.1093/nsr/nwac259} {\bibfield  {journal} {\bibinfo  {journal} {National Science Review}\ }\textbf {\bibinfo {volume} {10}},\ \bibinfo {pages} {nwac259} (\bibinfo {year} {2022})}\BibitemShut {NoStop}%
\bibitem [{\citenamefont {Peters}\ and\ \citenamefont {Rodriguez}(2022)}]{PhysRevLett.129.013901}%
  \BibitemOpen
  \bibfield  {author} {\bibinfo {author} {\bibfnamefont {K.~J.~H.}\ \bibnamefont {Peters}}\ and\ \bibinfo {author} {\bibfnamefont {S.~R.~K.}\ \bibnamefont {Rodriguez}},\ }\bibfield  {title} {\bibinfo {title} {Exceptional precision of a nonlinear optical sensor at a square-root singularity},\ }\href {https://doi.org/10.1103/PhysRevLett.129.013901} {\bibfield  {journal} {\bibinfo  {journal} {Phys. Rev. Lett.}\ }\textbf {\bibinfo {volume} {129}},\ \bibinfo {pages} {013901} (\bibinfo {year} {2022})}\BibitemShut {NoStop}%
\bibitem [{\citenamefont {Bai}\ \emph {et~al.}(2023)\citenamefont {Bai}, \citenamefont {Li}, \citenamefont {Liu}, \citenamefont {Fang}, \citenamefont {Wan},\ and\ \citenamefont {Xiao}}]{PhysRevLett.130.266901}%
  \BibitemOpen
  \bibfield  {author} {\bibinfo {author} {\bibfnamefont {K.}~\bibnamefont {Bai}}, \bibinfo {author} {\bibfnamefont {J.-Z.}\ \bibnamefont {Li}}, \bibinfo {author} {\bibfnamefont {T.-R.}\ \bibnamefont {Liu}}, \bibinfo {author} {\bibfnamefont {L.}~\bibnamefont {Fang}}, \bibinfo {author} {\bibfnamefont {D.}~\bibnamefont {Wan}},\ and\ \bibinfo {author} {\bibfnamefont {M.}~\bibnamefont {Xiao}},\ }\bibfield  {title} {\bibinfo {title} {Nonlinear exceptional points with a complete basis in dynamics},\ }\href {https://doi.org/10.1103/PhysRevLett.130.266901} {\bibfield  {journal} {\bibinfo  {journal} {Phys. Rev. Lett.}\ }\textbf {\bibinfo {volume} {130}},\ \bibinfo {pages} {266901} (\bibinfo {year} {2023})}\BibitemShut {NoStop}%
\bibitem [{\citenamefont {Roy}\ \emph {et~al.}(2021)\citenamefont {Roy}, \citenamefont {Jahani}, \citenamefont {Guo}, \citenamefont {Dutt}, \citenamefont {Fan}, \citenamefont {Miri},\ and\ \citenamefont {Marandi}}]{Roy:21}%
  \BibitemOpen
  \bibfield  {author} {\bibinfo {author} {\bibfnamefont {A.}~\bibnamefont {Roy}}, \bibinfo {author} {\bibfnamefont {S.}~\bibnamefont {Jahani}}, \bibinfo {author} {\bibfnamefont {Q.}~\bibnamefont {Guo}}, \bibinfo {author} {\bibfnamefont {A.}~\bibnamefont {Dutt}}, \bibinfo {author} {\bibfnamefont {S.}~\bibnamefont {Fan}}, \bibinfo {author} {\bibfnamefont {M.-A.}\ \bibnamefont {Miri}},\ and\ \bibinfo {author} {\bibfnamefont {A.}~\bibnamefont {Marandi}},\ }\bibfield  {title} {\bibinfo {title} {Nondissipative non-hermitian dynamics and exceptional points in coupled optical parametric oscillators},\ }\href {https://doi.org/10.1364/OPTICA.415569} {\bibfield  {journal} {\bibinfo  {journal} {Optica}\ }\textbf {\bibinfo {volume} {8}},\ \bibinfo {pages} {415} (\bibinfo {year} {2021})}\BibitemShut {NoStop}%
\bibitem [{\citenamefont {Longhi}\ and\ \citenamefont {Feng}(2017)}]{Longhi:17}%
  \BibitemOpen
  \bibfield  {author} {\bibinfo {author} {\bibfnamefont {S.}~\bibnamefont {Longhi}}\ and\ \bibinfo {author} {\bibfnamefont {L.}~\bibnamefont {Feng}},\ }\bibfield  {title} {\bibinfo {title} {Unidirectional lasing in semiconductor microring lasers at an exceptional point [invited]},\ }\href {https://doi.org/10.1364/PRJ.5.0000B1} {\bibfield  {journal} {\bibinfo  {journal} {Photon. Res.}\ }\textbf {\bibinfo {volume} {5}},\ \bibinfo {pages} {B1} (\bibinfo {year} {2017})}\BibitemShut {NoStop}%
\bibitem [{\citenamefont {Ji}\ \emph {et~al.}(2023)\citenamefont {Ji}, \citenamefont {Zhong}, \citenamefont {Ge}, \citenamefont {Beaudoin}, \citenamefont {Sagnes}, \citenamefont {Raineri}, \citenamefont {El-Ganainy},\ and\ \citenamefont {Yacomotti}}]{Ji2023}%
  \BibitemOpen
  \bibfield  {author} {\bibinfo {author} {\bibfnamefont {K.}~\bibnamefont {Ji}}, \bibinfo {author} {\bibfnamefont {Q.}~\bibnamefont {Zhong}}, \bibinfo {author} {\bibfnamefont {L.}~\bibnamefont {Ge}}, \bibinfo {author} {\bibfnamefont {G.}~\bibnamefont {Beaudoin}}, \bibinfo {author} {\bibfnamefont {I.}~\bibnamefont {Sagnes}}, \bibinfo {author} {\bibfnamefont {F.}~\bibnamefont {Raineri}}, \bibinfo {author} {\bibfnamefont {R.}~\bibnamefont {El-Ganainy}},\ and\ \bibinfo {author} {\bibfnamefont {A.~M.}\ \bibnamefont {Yacomotti}},\ }\bibfield  {title} {\bibinfo {title} {Tracking exceptional points above the lasing threshold},\ }\href {https://doi.org/10.1038/s41467-023-43874-z} {\bibfield  {journal} {\bibinfo  {journal} {Nature Communications}\ }\textbf {\bibinfo {volume} {14}},\ \bibinfo {pages} {8304} (\bibinfo {year} {2023})}\BibitemShut {NoStop}%
\bibitem [{\citenamefont {Fischer}\ \emph {et~al.}(2024)\citenamefont {Fischer}, \citenamefont {Raziman}, \citenamefont {Ng}, \citenamefont {Clarysse}, \citenamefont {Saxena}, \citenamefont {Dranczewski}, \citenamefont {Vezzoli}, \citenamefont {Schmid}, \citenamefont {Moselund},\ and\ \citenamefont {Sapienza}}]{Fischer2024}%
  \BibitemOpen
  \bibfield  {author} {\bibinfo {author} {\bibfnamefont {A.}~\bibnamefont {Fischer}}, \bibinfo {author} {\bibfnamefont {T.~V.}\ \bibnamefont {Raziman}}, \bibinfo {author} {\bibfnamefont {W.~K.}\ \bibnamefont {Ng}}, \bibinfo {author} {\bibfnamefont {J.}~\bibnamefont {Clarysse}}, \bibinfo {author} {\bibfnamefont {D.}~\bibnamefont {Saxena}}, \bibinfo {author} {\bibfnamefont {J.}~\bibnamefont {Dranczewski}}, \bibinfo {author} {\bibfnamefont {S.}~\bibnamefont {Vezzoli}}, \bibinfo {author} {\bibfnamefont {H.}~\bibnamefont {Schmid}}, \bibinfo {author} {\bibfnamefont {K.}~\bibnamefont {Moselund}},\ and\ \bibinfo {author} {\bibfnamefont {R.}~\bibnamefont {Sapienza}},\ }\bibfield  {title} {\bibinfo {title} {Controlling lasing around exceptional points in coupled nanolasers},\ }\href {https://doi.org/10.1038/s44310-024-00006-9} {\bibfield  {journal} {\bibinfo  {journal} {npj Nanophotonics}\ }\textbf {\bibinfo {volume} {1}},\ \bibinfo {pages} {6} (\bibinfo {year} {2024})}\BibitemShut {NoStop}%
\bibitem [{\citenamefont {Siegman}(1986)}]{siegman1986lasers}%
  \BibitemOpen
  \bibfield  {author} {\bibinfo {author} {\bibfnamefont {A.~E.}\ \bibnamefont {Siegman}},\ }\href@noop {} {\emph {\bibinfo {title} {Lasers}}}\ (\bibinfo  {publisher} {University science books},\ \bibinfo {year} {1986})\BibitemShut {NoStop}%
\bibitem [{\citenamefont {Lang}(1982)}]{1071632}%
  \BibitemOpen
  \bibfield  {author} {\bibinfo {author} {\bibfnamefont {R.}~\bibnamefont {Lang}},\ }\bibfield  {title} {\bibinfo {title} {Injection locking properties of a semiconductor laser},\ }\href {https://doi.org/10.1109/JQE.1982.1071632} {\bibfield  {journal} {\bibinfo  {journal} {IEEE Journal of Quantum Electronics}\ }\textbf {\bibinfo {volume} {18}},\ \bibinfo {pages} {976} (\bibinfo {year} {1982})}\BibitemShut {NoStop}%
\bibitem [{\citenamefont {Kobayashi}\ and\ \citenamefont {Kimura}(1981)}]{1071166}%
  \BibitemOpen
  \bibfield  {author} {\bibinfo {author} {\bibfnamefont {S.}~\bibnamefont {Kobayashi}}\ and\ \bibinfo {author} {\bibfnamefont {T.}~\bibnamefont {Kimura}},\ }\bibfield  {title} {\bibinfo {title} {Injection locking in algaas semiconductor laser},\ }\href {https://doi.org/10.1109/JQE.1981.1071166} {\bibfield  {journal} {\bibinfo  {journal} {IEEE Journal of Quantum Electronics}\ }\textbf {\bibinfo {volume} {17}},\ \bibinfo {pages} {681} (\bibinfo {year} {1981})}\BibitemShut {NoStop}%
\bibitem [{\citenamefont {Lang}\ and\ \citenamefont {Kobayashi}(1980)}]{1070479}%
  \BibitemOpen
  \bibfield  {author} {\bibinfo {author} {\bibfnamefont {R.}~\bibnamefont {Lang}}\ and\ \bibinfo {author} {\bibfnamefont {K.}~\bibnamefont {Kobayashi}},\ }\bibfield  {title} {\bibinfo {title} {External optical feedback effects on semiconductor injection laser properties},\ }\href {https://doi.org/10.1109/JQE.1980.1070479} {\bibfield  {journal} {\bibinfo  {journal} {IEEE Journal of Quantum Electronics}\ }\textbf {\bibinfo {volume} {16}},\ \bibinfo {pages} {347} (\bibinfo {year} {1980})}\BibitemShut {NoStop}%
\bibitem [{\citenamefont {Noda}\ \emph {et~al.}(1990)\citenamefont {Noda}, \citenamefont {Kojima},\ and\ \citenamefont {Kyuma}}]{62107}%
  \BibitemOpen
  \bibfield  {author} {\bibinfo {author} {\bibfnamefont {S.}~\bibnamefont {Noda}}, \bibinfo {author} {\bibfnamefont {K.}~\bibnamefont {Kojima}},\ and\ \bibinfo {author} {\bibfnamefont {K.}~\bibnamefont {Kyuma}},\ }\bibfield  {title} {\bibinfo {title} {Mutual injection-locking properties of monolithically-integrated surface-emitting multiple-quantum-well distributed feedback lasers},\ }\href {https://doi.org/10.1109/3.62107} {\bibfield  {journal} {\bibinfo  {journal} {IEEE Journal of Quantum Electronics}\ }\textbf {\bibinfo {volume} {26}},\ \bibinfo {pages} {1883} (\bibinfo {year} {1990})}\BibitemShut {NoStop}%
\bibitem [{\citenamefont {Takata}\ \emph {et~al.}(2015)\citenamefont {Takata}, \citenamefont {Marandi},\ and\ \citenamefont {Yamamoto}}]{PhysRevA.92.043821}%
  \BibitemOpen
  \bibfield  {author} {\bibinfo {author} {\bibfnamefont {K.}~\bibnamefont {Takata}}, \bibinfo {author} {\bibfnamefont {A.}~\bibnamefont {Marandi}},\ and\ \bibinfo {author} {\bibfnamefont {Y.}~\bibnamefont {Yamamoto}},\ }\bibfield  {title} {\bibinfo {title} {Quantum correlation in degenerate optical parametric oscillators with mutual injections},\ }\href {https://doi.org/10.1103/PhysRevA.92.043821} {\bibfield  {journal} {\bibinfo  {journal} {Phys. Rev. A}\ }\textbf {\bibinfo {volume} {92}},\ \bibinfo {pages} {043821} (\bibinfo {year} {2015})}\BibitemShut {NoStop}%
\bibitem [{\citenamefont {Wang}\ \emph {et~al.}(2013)\citenamefont {Wang}, \citenamefont {Marandi}, \citenamefont {Wen}, \citenamefont {Byer},\ and\ \citenamefont {Yamamoto}}]{PhysRevA.88.063853}%
  \BibitemOpen
  \bibfield  {author} {\bibinfo {author} {\bibfnamefont {Z.}~\bibnamefont {Wang}}, \bibinfo {author} {\bibfnamefont {A.}~\bibnamefont {Marandi}}, \bibinfo {author} {\bibfnamefont {K.}~\bibnamefont {Wen}}, \bibinfo {author} {\bibfnamefont {R.~L.}\ \bibnamefont {Byer}},\ and\ \bibinfo {author} {\bibfnamefont {Y.}~\bibnamefont {Yamamoto}},\ }\bibfield  {title} {\bibinfo {title} {Coherent ising machine based on degenerate optical parametric oscillators},\ }\href {https://doi.org/10.1103/PhysRevA.88.063853} {\bibfield  {journal} {\bibinfo  {journal} {Phys. Rev. A}\ }\textbf {\bibinfo {volume} {88}},\ \bibinfo {pages} {063853} (\bibinfo {year} {2013})}\BibitemShut {NoStop}%
\bibitem [{\citenamefont {Inagaki}\ \emph {et~al.}(2016)\citenamefont {Inagaki}, \citenamefont {Haribara}, \citenamefont {Igarashi}, \citenamefont {Sonobe}, \citenamefont {Tamate}, \citenamefont {Honjo}, \citenamefont {Marandi}, \citenamefont {McMahon}, \citenamefont {Umeki}, \citenamefont {Enbutsu}, \citenamefont {Tadanaga}, \citenamefont {Takenouchi}, \citenamefont {Aihara}, \citenamefont {ichi Kawarabayashi}, \citenamefont {Inoue}, \citenamefont {Utsunomiya},\ and\ \citenamefont {Takesue}}]{doi:10.1126/science.aah4243}%
  \BibitemOpen
  \bibfield  {author} {\bibinfo {author} {\bibfnamefont {T.}~\bibnamefont {Inagaki}}, \bibinfo {author} {\bibfnamefont {Y.}~\bibnamefont {Haribara}}, \bibinfo {author} {\bibfnamefont {K.}~\bibnamefont {Igarashi}}, \bibinfo {author} {\bibfnamefont {T.}~\bibnamefont {Sonobe}}, \bibinfo {author} {\bibfnamefont {S.}~\bibnamefont {Tamate}}, \bibinfo {author} {\bibfnamefont {T.}~\bibnamefont {Honjo}}, \bibinfo {author} {\bibfnamefont {A.}~\bibnamefont {Marandi}}, \bibinfo {author} {\bibfnamefont {P.~L.}\ \bibnamefont {McMahon}}, \bibinfo {author} {\bibfnamefont {T.}~\bibnamefont {Umeki}}, \bibinfo {author} {\bibfnamefont {K.}~\bibnamefont {Enbutsu}}, \bibinfo {author} {\bibfnamefont {O.}~\bibnamefont {Tadanaga}}, \bibinfo {author} {\bibfnamefont {H.}~\bibnamefont {Takenouchi}}, \bibinfo {author} {\bibfnamefont {K.}~\bibnamefont {Aihara}}, \bibinfo {author} {\bibfnamefont {K.}~\bibnamefont {ichi Kawarabayashi}}, \bibinfo {author} {\bibfnamefont {K.}~\bibnamefont {Inoue}}, \bibinfo {author} {\bibfnamefont {S.}~\bibnamefont {Utsunomiya}},\ and\ \bibinfo {author} {\bibfnamefont {H.}~\bibnamefont {Takesue}},\ }\bibfield  {title} {\bibinfo {title} {A coherent ising machine for 2000-node optimization problems},\ }\href {https://doi.org/10.1126/science.aah4243} {\bibfield  {journal} {\bibinfo  {journal} {Science}\ }\textbf {\bibinfo {volume} {354}},\ \bibinfo {pages} {603} (\bibinfo {year} {2016})}\BibitemShut {NoStop}%
\bibitem [{\citenamefont {Takemura}\ \emph {et~al.}(2021)\citenamefont {Takemura}, \citenamefont {Takata}, \citenamefont {Takiguchi},\ and\ \citenamefont {Notomi}}]{Takemura2021}%
  \BibitemOpen
  \bibfield  {author} {\bibinfo {author} {\bibfnamefont {N.}~\bibnamefont {Takemura}}, \bibinfo {author} {\bibfnamefont {K.}~\bibnamefont {Takata}}, \bibinfo {author} {\bibfnamefont {M.}~\bibnamefont {Takiguchi}},\ and\ \bibinfo {author} {\bibfnamefont {M.}~\bibnamefont {Notomi}},\ }\bibfield  {title} {\bibinfo {title} {Emulating the local kuramoto model with an injection-locked photonic crystal laser array},\ }\href {https://doi.org/10.1038/s41598-021-86982-w} {\bibfield  {journal} {\bibinfo  {journal} {Scientific Reports}\ }\textbf {\bibinfo {volume} {11}},\ \bibinfo {pages} {8587} (\bibinfo {year} {2021})}\BibitemShut {NoStop}%
\bibitem [{\citenamefont {Madiot}\ \emph {et~al.}(2024)\citenamefont {Madiot}, \citenamefont {Chateiller}, \citenamefont {Bazin}, \citenamefont {Loren}, \citenamefont {Pantzas}, \citenamefont {Beaudoin}, \citenamefont {Sagnes},\ and\ \citenamefont {Raineri}}]{doi:10.1126/sciadv.adr8283}%
  \BibitemOpen
  \bibfield  {author} {\bibinfo {author} {\bibfnamefont {G.}~\bibnamefont {Madiot}}, \bibinfo {author} {\bibfnamefont {Q.}~\bibnamefont {Chateiller}}, \bibinfo {author} {\bibfnamefont {A.}~\bibnamefont {Bazin}}, \bibinfo {author} {\bibfnamefont {P.}~\bibnamefont {Loren}}, \bibinfo {author} {\bibfnamefont {K.}~\bibnamefont {Pantzas}}, \bibinfo {author} {\bibfnamefont {G.}~\bibnamefont {Beaudoin}}, \bibinfo {author} {\bibfnamefont {I.}~\bibnamefont {Sagnes}},\ and\ \bibinfo {author} {\bibfnamefont {F.}~\bibnamefont {Raineri}},\ }\bibfield  {title} {\bibinfo {title} {Harnessing coupled nanolasers near exceptional points for directional emission},\ }\href {https://doi.org/10.1126/sciadv.adr8283} {\bibfield  {journal} {\bibinfo  {journal} {Science Advances}\ }\textbf {\bibinfo {volume} {10}},\ \bibinfo {pages} {eadr8283} (\bibinfo {year} {2024})}\BibitemShut {NoStop}%
\bibitem [{\citenamefont {and}(2016)}]{Nakao02042016}%
  \BibitemOpen
  \bibfield  {author} {\bibinfo {author} {\bibfnamefont {H.~N.}\ \bibnamefont {and}},\ }\bibfield  {title} {\bibinfo {title} {Phase reduction approach to synchronisation of nonlinear oscillators},\ }\href {https://doi.org/10.1080/00107514.2015.1094987} {\bibfield  {journal} {\bibinfo  {journal} {Contemporary Physics}\ }\textbf {\bibinfo {volume} {57}},\ \bibinfo {pages} {188} (\bibinfo {year} {2016})}\BibitemShut {NoStop}%
\bibitem [{\citenamefont {Mulet}\ \emph {et~al.}(2002)\citenamefont {Mulet}, \citenamefont {Masoller},\ and\ \citenamefont {Mirasso}}]{PhysRevA.65.063815}%
  \BibitemOpen
  \bibfield  {author} {\bibinfo {author} {\bibfnamefont {J.}~\bibnamefont {Mulet}}, \bibinfo {author} {\bibfnamefont {C.}~\bibnamefont {Masoller}},\ and\ \bibinfo {author} {\bibfnamefont {C.~R.}\ \bibnamefont {Mirasso}},\ }\bibfield  {title} {\bibinfo {title} {Modeling bidirectionally coupled single-mode semiconductor lasers},\ }\href {https://doi.org/10.1103/PhysRevA.65.063815} {\bibfield  {journal} {\bibinfo  {journal} {Phys. Rev. A}\ }\textbf {\bibinfo {volume} {65}},\ \bibinfo {pages} {063815} (\bibinfo {year} {2002})}\BibitemShut {NoStop}%
\bibitem [{\citenamefont {Winful}\ and\ \citenamefont {Rahman}(1990)}]{PhysRevLett.65.1575}%
  \BibitemOpen
  \bibfield  {author} {\bibinfo {author} {\bibfnamefont {H.~G.}\ \bibnamefont {Winful}}\ and\ \bibinfo {author} {\bibfnamefont {L.}~\bibnamefont {Rahman}},\ }\bibfield  {title} {\bibinfo {title} {Synchronized chaos and spatiotemporal chaos in arrays of coupled lasers},\ }\href {https://doi.org/10.1103/PhysRevLett.65.1575} {\bibfield  {journal} {\bibinfo  {journal} {Phys. Rev. Lett.}\ }\textbf {\bibinfo {volume} {65}},\ \bibinfo {pages} {1575} (\bibinfo {year} {1990})}\BibitemShut {NoStop}%
\bibitem [{\citenamefont {Thornburg}\ \emph {et~al.}(1997)\citenamefont {Thornburg}, \citenamefont {M\"oller}, \citenamefont {Roy}, \citenamefont {Carr}, \citenamefont {Li},\ and\ \citenamefont {Erneux}}]{PhysRevE.55.3865}%
  \BibitemOpen
  \bibfield  {author} {\bibinfo {author} {\bibfnamefont {K.~S.}\ \bibnamefont {Thornburg}}, \bibinfo {author} {\bibfnamefont {M.}~\bibnamefont {M\"oller}}, \bibinfo {author} {\bibfnamefont {R.}~\bibnamefont {Roy}}, \bibinfo {author} {\bibfnamefont {T.~W.}\ \bibnamefont {Carr}}, \bibinfo {author} {\bibfnamefont {R.-D.}\ \bibnamefont {Li}},\ and\ \bibinfo {author} {\bibfnamefont {T.}~\bibnamefont {Erneux}},\ }\bibfield  {title} {\bibinfo {title} {Chaos and coherence in coupled lasers},\ }\href {https://doi.org/10.1103/PhysRevE.55.3865} {\bibfield  {journal} {\bibinfo  {journal} {Phys. Rev. E}\ }\textbf {\bibinfo {volume} {55}},\ \bibinfo {pages} {3865} (\bibinfo {year} {1997})}\BibitemShut {NoStop}%
\bibitem [{\citenamefont {Naweed}(2015)}]{Naweed:15}%
  \BibitemOpen
  \bibfield  {author} {\bibinfo {author} {\bibfnamefont {A.}~\bibnamefont {Naweed}},\ }\bibfield  {title} {\bibinfo {title} {Photonic coherence effects from dual-waveguide coupled pair of co-resonant microring resonators},\ }\href {https://doi.org/10.1364/OE.23.012573} {\bibfield  {journal} {\bibinfo  {journal} {Opt. Express}\ }\textbf {\bibinfo {volume} {23}},\ \bibinfo {pages} {12573} (\bibinfo {year} {2015})}\BibitemShut {NoStop}%
\bibitem [{\citenamefont {Wang}\ \emph {et~al.}(2020)\citenamefont {Wang}, \citenamefont {Jiang}, \citenamefont {Zhao}, \citenamefont {Zhang}, \citenamefont {Hsu}, \citenamefont {Peng}, \citenamefont {Stone}, \citenamefont {Jiang},\ and\ \citenamefont {Yang}}]{Wang2020}%
  \BibitemOpen
  \bibfield  {author} {\bibinfo {author} {\bibfnamefont {C.}~\bibnamefont {Wang}}, \bibinfo {author} {\bibfnamefont {X.}~\bibnamefont {Jiang}}, \bibinfo {author} {\bibfnamefont {G.}~\bibnamefont {Zhao}}, \bibinfo {author} {\bibfnamefont {M.}~\bibnamefont {Zhang}}, \bibinfo {author} {\bibfnamefont {C.~W.}\ \bibnamefont {Hsu}}, \bibinfo {author} {\bibfnamefont {B.}~\bibnamefont {Peng}}, \bibinfo {author} {\bibfnamefont {A.~D.}\ \bibnamefont {Stone}}, \bibinfo {author} {\bibfnamefont {L.}~\bibnamefont {Jiang}},\ and\ \bibinfo {author} {\bibfnamefont {L.}~\bibnamefont {Yang}},\ }\bibfield  {title} {\bibinfo {title} {Electromagnetically induced transparency at a chiral exceptional point},\ }\href {https://doi.org/10.1038/s41567-019-0746-7} {\bibfield  {journal} {\bibinfo  {journal} {Nature Physics}\ }\textbf {\bibinfo {volume} {16}},\ \bibinfo {pages} {334} (\bibinfo {year} {2020})}\BibitemShut {NoStop}%
\bibitem [{\citenamefont {Sakaguchi}\ \emph {et~al.}(1988)\citenamefont {Sakaguchi}, \citenamefont {Shinomoto},\ and\ \citenamefont {Kuramoto}}]{10.1143/PTP.79.1069}%
  \BibitemOpen
  \bibfield  {author} {\bibinfo {author} {\bibfnamefont {H.}~\bibnamefont {Sakaguchi}}, \bibinfo {author} {\bibfnamefont {S.}~\bibnamefont {Shinomoto}},\ and\ \bibinfo {author} {\bibfnamefont {Y.}~\bibnamefont {Kuramoto}},\ }\bibfield  {title} {\bibinfo {title} {Mutual entrainment in oscillator lattices with nonvariational type interaction},\ }\href {https://doi.org/10.1143/PTP.79.1069} {\bibfield  {journal} {\bibinfo  {journal} {Progress of Theoretical Physics}\ }\textbf {\bibinfo {volume} {79}},\ \bibinfo {pages} {1069} (\bibinfo {year} {1988})}\BibitemShut {NoStop}%
\end{thebibliography}%

\end{document}